\documentclass[12pt, letterpaper]{article}
\usepackage{amsmath}
\usepackage{amsfonts}
\usepackage{amssymb}
\usepackage{graphicx}
\usepackage{caption}
\usepackage{subcaption}
\usepackage{graphics,epsfig}
\usepackage{epsf}
\usepackage{epstopdf}
\usepackage{color}
\usepackage{float}
\numberwithin{equation}{section}

\newcommand{\nc}{\newcommand}
\nc{\ba}{\begin{eqnarray}}
\nc{\ea}{\end{eqnarray}}
\newcommand\be{\begin{equation}}
\newcommand\ee{\end{equation}}
\newcommand{\calR}{{\cal{R}}}
\newcommand{\calP}{{\cal{P}}}
\nc{\e}{{\bf{e}}}
\nc{\kk}{{\bf{k}}}
\nc{\pp}{{\bf{p}}}
\nc{\cM}{{\cal{M}}}

\begin{document}

\vspace{5mm}
\vspace{0.5cm}
\begin{center}

\def\thefootnote{\fnsymbol{footnote}}

{\Large {\bf  Charged Vector Inflation}}
\\[1cm]

{Hassan Firouzjahi$^{1}\footnote{firouz@ipm.ir }$, Mohammad Ali Gorji$^{1}\footnote{gorji@ipm.ir}$,    
Seyed Ali Hosseini Mansoori$^{ 2}\footnote{shosseini@shahroodut.ac.ir}$, \\
Asieh Karami$^{1}\footnote{karami@ipm.ir}$, Tahereh Rostami$^{1}\footnote{t.rostami@ipm.ir}$,}
\\[0.5cm]

{\small \textit{$^1$School of Astronomy, 
Institute for Research in Fundamental Sciences (IPM) \\ 
P.~O.~Box 19395-5531, Tehran, Iran }} \\
\vspace{0.5cm}

{\small \textit{$^2$Faculty of Physics, Shahrood University of Technology, P.O. Box 3619995161, Shahrood, Iran }}

\end{center}

\vspace{.8cm}

\begin{abstract}
We present a model of inflation in which the inflaton field is charged under a triplet of $U(1)$ gauge fields. The model enjoys an internal $O(3)$ symmetry  supporting the isotropic FRW solution. With an appropriate coupling between the gauge fields and the inflaton field, the system reaches an attractor regime in which the gauge fields furnish a small constant fraction of the total energy density. We decompose the scalar perturbations into the adiabatic and entropy modes and calculate the contributions of the gauge fields into the curvature perturbations power spectrum. We also calculate the entropy power spectrum and the adiabatic-entropy cross correlation. In addition to the metric tensor perturbations, there are tensor perturbations associated with the gauge field perturbations which are coupled to metric tensor perturbations. We show that the correction in primordial gravitational  tensor power spectrum induced from the matter tensor perturbation  is a sensitive function of the gauge coupling. 

\end{abstract}
\vspace{0.5cm} \hrule
\def\thefootnote{\arabic{footnote}}
\setcounter{footnote}{0}

\newpage

\section{Introduction}

Models of inflation based on a single scalar field with a flat potential are well consistent with cosmological observations \cite{Akrami:2018odb, Ade:2015lrj}. Among the basic predictions of models of inflation are that the primordial perturbations are nearly scale invariant, nearly adiabatic and nearly Gaussian, in very good agreements with observations. Having said this, there is no unique realization of inflation dynamics in the context of high energy physics or  beyond Standard Model (SM) of particle physics.  
For example, what is the nature of the inflaton field(s)? What mechanism keeps the inflationary potential flat enough to sustain a long enough period of inflation to solve the flatness and the horizon problems?

It is generally believed that there may exist many fields during  inflation which can play some roles. If the fields are very heavy compared to the Hubble scale during inflation, then they are not expected to play important roles. However, if the fields are light or semi-heavy they can have non-trivial effects on cosmological observables such as the power spectrum and bispectrum, see for example 
\cite{ Chen:2009zp, Noumi:2012vr,  Emami:2013lma}.  In addition, there is no reason that only scalar fields play important roles during inflation. Specifically, the gauge fields and vector fields are essential ingredients of SM and any theory of high energy physics. Therefore, it is quite natural to look for the imprints of the vector fields during inflation. One issue with the vector fields in background is that they have preferred  directions so in general models of inflation with background vector fields are anisotropic. The second issue with the vector fields is that because of the conformal invariance,  they are quickly diluted in an expanding background, so their effects  become rapidly insignificant during inflation.  

Anisotropic inflation is a model of inflation based on a $U(1)$ gauge field dynamics. To remedy the second issue mentioned above, the gauge kinetic coupling in these models is a function of the inflaton field so the conformal invariance is broken. By choosing 
an appropriate form of the gauge kinetic coupling, the electric field energy density becomes nearly constant so the gauge field survives the expansion till end of inflation \cite{Watanabe:2009ct}. In addition, the gauge field perturbations become nearly scale invariant and can take parts in generating cosmological perturbations. In particular, quadrupolar statistical anisotropies are generated in these models which can be tested in CMB maps. For various works on anisotropic inflation and their cosmological imprints see \cite{anisotropic-inflation}.

The anisotropic inflation model \cite{Watanabe:2009ct} has been extended to the
case where the scalar field is charged under the $U(1)$ gauge field in 
\cite{Emami1,Emami:2013bk,Chen:2014eua} while its isotropic realization containing a triplet of $U(1)$ gauge fields has been studied  in \cite{Yamamoto:2012sq, Funakoshi:2012ym} . In this work we consider the isotropic extension of \cite{Watanabe:2009ct} in which the inflaton is charged under a triplet of $U(1)$ gauge fields.  We show that the model has some interesting features such as it contains entropy 
mode in addition to the adiabatic mode and the gravitational tensor modes are sourced by the tensor modes coming from the gauge fields. 

The rest of the paper is organized as follows. In Section \ref{model} we present our setup and study its background dynamics.
In Section \ref{perturbations} we study the cosmological perturbations in this setup while the power spectra of the adiabatic and entropy perturbations and their cross correlations are studied in  Section \ref{sec-scalar}. The tensor perturbations of the metric and the matter fields are studied in Section \ref{sec-tensor} followed by the summaries and discussions  in Section \ref{summary}.  The  gauge symmetries of the setup
are studied in \ref{gauge-symmetries} while the analysis of quadratic action are relegated into the Appendix \ref{App-action}.


\section{The Setup and Background Dynamics}
\label{model}

In this Section we introduce our setup in which we  extend the model of anisotropic inflation to the setup which can support isotropic FRW solution. A realization of this was studied in  \cite{Yamamoto:2012sq, Funakoshi:2012ym} in which the model contains a triplet of $U(1)$ gauge fields with an additional global internal $O(3)$ symmetry. The internal $O(3)$ symmetry allows one to obtain isotropic FRW solution \cite{Emami:2016ldl}. In this work, we extend the setup of  \cite{Yamamoto:2012sq, Funakoshi:2012ym}
to a model  containing three complex scalar fields $\phi_{(a)}, a=1,2, 3$, 
charged under $U(1)_a$ gauge symmetry with gauge coupling $\e$.  In a sense our setup is the isotropic realization of the model of anisotropic charged inflation studied in \cite{Emami1,Emami:2013bk,Chen:2014eua}.

\subsection{The Setup}

We consider a model consisting of a triplet of $U(1)$ gauge fields which may be thought as three independent copies of the  $U(1)$ scalar electrodynamics. The desired gauge symmetry is $U(1)_a=U(1)_1\times U(1)_2\times U(1)_3$ and the scalar sector is defined by a triplet $\Phi$ 
\be\label{Phi}
\Phi = \begin{pmatrix} \phi_{(1)} \\ \phi_{(2)} \\ \phi_{(3)} \end{pmatrix} \,,
\ee
in which $\phi_{(a)}, a=1,2, 3$ are complex scalar fields which are charged under $U(1)_a$ the gauge field and couple to the gauge fields ${\mathbf A}_{\mu}$ through the covariant derivative denoted as

\begin{eqnarray}\label{CD}
{\mathbf D}_{\mu} = {\mathbf 1} \partial_{\mu} + i {\e} {\mathbf A}_{\mu} \,.
\end{eqnarray}
The gauge coupling constant $\e$ assigns the same charges to each scalar field. 

Similar to the original model of anisotropic inflation \cite{Watanabe:2009ct}, the action of the model is given by 

\begin{equation}\label{action01}
S = \int d^{4}x \sqrt{-g} \Big[ \frac{M_P^2}{2}R  
- \frac{1}{2} ({\mathbf D}_{\mu} \Phi)^\dagger ({\mathbf D}^{\mu} \Phi) - V(|\Phi|)
- \frac{1}{4} f^{2}(|\Phi|) \,
\mbox{Tr}\big({\mathbf F}_{\mu \nu} {\mathbf F}^{\mu \nu}\big) \Big] \,,
\end{equation}
where $M_P$ is the reduced Planck mass, $R$ is the Ricci scalar,
$|\Phi|=\sqrt{\Phi^\dagger\Phi}$, $V$ is the potential, $f$ is the conformal factor 
and ${\mathbf F}_{\mu\nu}$
is the field strength tensor defined in the spirit of the covariant derivative 
(\ref{CD}). To simplify the setup, we have assumed that  $V$ and $f$ are only functions of the magnitude $ |\Phi|$.

The details of  the gauge symmetries  of the model are presented in Appendix \ref{gauge-symmetries}. Gauge fields $A_\mu^{(a)}$ enjoy the associated $U(1)_a$ gauge symmetry for $a=1, 2$ and $3$. To fix the $U(1)_a$ gauge freedoms, we work in the gauge where all scalar fields $\phi_{(a)}$ are real. In other words, we fix the $U(1)_a$ gauges by going to unitary gauge where the phases of the complex scalar field are set to zero.
In addition, in order to obtain isotropic FRW solution, similar to the setup of  \cite{Papadopoulos:2017xxx}, we consider a subset of the model in which 
$\phi_{(1)} = \phi_{(2)} = \phi_{(3)} \equiv  \phi/\sqrt{3}$ where the kinetic term $ ({\mathbf D}_{\mu} \Phi)^\dagger ({\mathbf D}^{\mu} \Phi)$ takes the
isotropic  form
\begin{eqnarray}
\label{kinetic-o3}
({\mathbf D}_{\mu}\Phi)^\dagger ({\mathbf D}^{\mu} \Phi) = 
\partial_{\mu}\phi \partial^{\mu}\phi 
+ \frac{\e^2}{3}  \phi^2 A_{\mu}^{(a) } A_{(a) }^{\mu} \,.
\end{eqnarray}

Putting these all together,   the action (\ref{action01}) takes the following isotropic form
\begin{equation}\label{action}
S = \int d^{4}x \sqrt{-g} \Big[ \frac{M_P^2}{2}R  
- \frac{1}{2} \partial_{\mu}\phi \partial^{\mu}\phi 
- \frac{\e^2}{6}  \phi^2 A_{\mu}^{(a)} A_{(a)}^{\mu} - V(\phi)
- \frac{1}{4} f^{2}(\phi) \, F^{(a)}_{\mu \nu} F_{(a)}^{\mu \nu} \Big] \,.
\end{equation}
As expected, the action (\ref{action}) has the same form as in models of anisotropic inflation \cite{Watanabe:2009ct} but  the gauge fields here enjoy an additional internal $O(3)$ symmetry, admitting FRW background solution. As in  \cite{Watanabe:2009ct} the conformal coupling $f(\phi)$ will be chosen such that to prevent the dilution of the gauge field energy density in the inflationary background. 
 
It is constructive to  compare our model with the other inflationary models that are constructed by means of $U(1)$ gauge fields. The  isotropic extension of the setup of anisotropic inflation\cite{Watanabe:2009ct} is suggested in \cite{Yamamoto:2012sq, Funakoshi:2012ym}  by means of a triplet of $U(1)$ gauge fields while the charged extension of  \cite{Watanabe:2009ct} is considered in \cite{Emami1}.  The model considered in \cite{Yamamoto:2012sq, Funakoshi:2012ym} has local $U(1)_a$ symmetry while it enjoys global $O(3)$ symmetry.  In this work we  have constructed the charged isotropic extension of anisotropic inflation \cite{Watanabe:2009ct}. In other words, our model is the charged generalization of \cite{Yamamoto:2012sq, Funakoshi:2012ym} and isotropic extension of \cite{Emami1}.

With the above discussions in mind, our setup with the action (\ref{action}) has similarities with the model studied in  \cite{Murata:2011wv} where the authors extended the setup of anisotropic inflation to a model where the inflaton field is coupled to a $SU(2)$ gauge kinetic function. In a sense the model considered in \cite{Murata:2011wv} can be thought as the charged extension of \cite{Yamamoto:2012sq, Funakoshi:2012ym}. 
The authors in \cite{Murata:2011wv} studied the background dynamics, verifying  the existence of the attractor solution and studying the shapes of anisotropies.  

It is worth mentioning that we can achieve the isotropic setup with more than two gauge fields \cite{Yamamoto:2012tq}, so having three gauge fields is the minimal setup which we have considered in this paper. Moreover, as was mentioned above, this case can be thought as the global limit of non-abelian gauge field models \cite{Murata:2011wv}.

\subsection{Background Equations }
\label{background}

Since  the action (\ref{action}) is $O(3)$ invariant, the model  admits the flat FRW cosmological background  
\begin{eqnarray}\label{FRW}
ds^2= - dt^2+ a(t)^2 \delta_{ij} dx^i dx^j \,,
\end{eqnarray}
with the ansatz \cite{Emami:2016ldl}
\ba
\label{3-A}
A_{\mu}^{(a)}(t) = A(t)\, \delta^{a}_{\mu} \, .
\ea
The model behaves like three mutually orthogonal gauge fields with $U(1)_a$ gauge symmetry and ansatz (\ref{3-A}) assigns the same magnitudes $A(t)$ to each gauge field 
\cite{Golovnev:2008cf}. Note that the ansatz (\ref{3-A}) is not the only solution. Indeed, one can imagine a situation in which the initial amplitudes of the gauge fields are not equal to each other, $A_{\mu}^{(a)}(t) \neq A_{\mu}^{(b)}(t)$ for $a \neq b$. In this case, the spacetime metric will be in the form of Bianchi type I Universe. However, as shown in \cite{Contaldi:2014zua}, one expects the isotropic FRW background to be the attractor solution of the system so the spacetime rapidly approaches the FRW background and the gauge field amplitudes become equal.  In addition, it is shown in \cite{Yamamoto:2012tq}, see also \cite{Golovnev:2008cf},  that with a large multiplet  of $U(1)$ gauge fields and with appropriate form of the conformal factor $f(\phi)$, the FRW solution is the attractor limit of arbitrary initial conditions with background anisotropies.  

Varying the action (\ref{action}) with respect to the gauge fields, we obtain the associated Maxwell equation
\begin{equation}\label{vector0}
\partial_t{\big( f^2a \dot{A} \big)}= - \frac{1}{3}\e^2 \phi^2 a A,
\end{equation}
where a dot indicates derivative with respect to the cosmic time $t$.

The variation of the action (\ref{action}) with respect to  the scalar field gives the Klein-Gordon equation
\begin{equation}
\label{q2}
\ddot\phi+3H\dot \phi+ V_{,\phi}=  \Big(
3ff_{,\phi}\dot{A}^2 -{1\over 3}\e^2 \phi A^2 \Big) a^{-2} \,,
\end{equation}
where $_{,\phi}$ denotes the derivative with respect to the scalar field. Note the important effects of the gauge field back-reactions on the scalar field as captured by the source term in the right hand side of the above equation. 

Finally, the corresponding Einstein equations are
\begin{eqnarray}
3 M_P^2 H^2 & = &\frac{1}{2}\dot
\phi^2+V(\phi)+ \frac{3f^2\dot{A}^2}{2a^2}
+\frac{\e^2\phi^2A^2}{6a^2} \label{q1} \,, \\
M_P^2 \big( 2\dot{H} + 3 H^2 \big)  & = & - \bigg( \frac{\dot{\phi}^2}{2} - V 
+\frac{f^2 \dot{A}^2}{2 a^2}-\frac{\e^2 A^2\phi^2}{6 a^2} \bigg)  \label{qq1} \,.
\end{eqnarray}
The right hand side of Eq. (\ref{q1}) is the total energy density while the  expression in the 
parentheses on the right hand side  of Eq. (\ref{qq1}) is the total pressure. In the absence 
of $\e$, from the above relations we see that the pure gauge fields contributions behaves 
like radiation thanks to the conformal symmetry. Let us consider the effects of the gauge 
coupling $\e$. We see from the second term in the right hand side of  Eq. (\ref{q2}) that the interaction 
$\e^2 \phi^2 A^2a^{-2}$  induces a time-dependent mass for the inflaton. However, the 
exponential time-dependence of this induced mass makes its main effect to occur towards 
the end of inflation where the exponential growth of the gauge field has its main influence.
Thus, to have a  long enough period of inflation, the back-reaction $\e^2 \phi^2 A^{\mu}A_{\mu}$ is negligible during much of the period of inflation and it only controls the mechanism of end of inflation 
\cite{Emami1, Emami:2013bk}. In this approximation one can easily solve the Maxwell 
equation (\ref{vector0}) to obtain

\begin{equation}\label{vector}
\dot{A}=\frac{q_{0}}{a} f^{-2} \, ,
\end{equation}
where $q_{0}$ is an integration constant. 

Now, as in anisotropic inflation model \cite{Watanabe:2009ct},  it is convenient to define the ratio of the energy density of the gauge fields to that of the inflaton as 
\begin{equation}
\label{R-eq}
R \equiv  \frac{\rho_{A}}{\rho_{\phi}}=\frac{3 q_{0}^2}{2V+\dot{\phi}^2} a^{-4}f^{-2} \, .
\end{equation}
In order to obtain a long period of inflation with a dS-like background, we expect that the contribution of the gauge field to the total energy density to be small. This is because, as just mentioned above, the gauge fields' contributions are like radiation and cannot support inflation by themselves. In other words, as in conventional models of slow-roll inflation, we expect that inflation to be driven  predominantly by the scalar field. As a result, we require $R \ll 1$ in order to obtain a long period of inflation. 

The dynamics of the background is very similar to the setup of anisotropic inflation.  During the early stage of inflation, the gauge fields do not drag enough energy from the inflaton field so the parameter $R$ is much smaller than the slow-roll parameters. In this limit, we can safely neglect the contributions of the gauge fields in total energy density and pressure and solve the system as in single field slow-roll models with 
\begin{equation}\label{eq3}
3 M_P^2 H^2 \simeq V \, , \quad \quad 
3 H \dot{\phi} \simeq -V_{,\phi}  \, .
\end{equation} 
Therefore, in the slow-roll limit, and for a given potential $V(\phi)$, the above equations provide the solution
\begin{equation}
a \simeq \exp\Big(-\frac{1}{M_P^2}\int_{\phi_{i}}^{\phi} \frac{V}{V_{,\phi}}d \phi\Big) \,.
\end{equation}
Now, as inflation proceeds, the gauge fields drag more and more energy from the inflaton field via the conformal coupling $f(\phi)$. As shown in \cite{Watanabe:2009ct} the system reaches an attractor limit in which the fraction of the gauge field energy density to total energy density reaches a constant value. During the attractor stage $R$ becomes at the order of slow-roll parameter and it stays nearly constant till end of inflation.  

In order for $R$ reach a constant value, from Eq. (\ref{R-eq}) one must choose $f(\phi) \propto a(t)^{-2}$. Therefore, it is reasonable to assume that
\begin{equation}\label{a1}
f(\phi)= \exp\Big(\frac{2 c}{M_P^2} \int \frac{V}{V_{,\phi}} d\phi\Big) \,,
\end{equation} 
with a constant parameter $c$.  

As the roles of the gauge fields become important, they back-react on the inflaton dynamics as given by the source term in Eq. (\ref{q2}). Taking into account 
the back-reactions of the gauge fields on the inflationary trajectory fix the relation between $R$ and the slow-roll parameter $\epsilon$. 

The scalar field equation in the slow-roll limit is given by 
\begin{equation}\label{a2}
3H\dot{\phi} = \frac{3 q_{0}^2 f,_{\phi}}{a^4 f^3} -V_{,\phi} \, .
\end{equation} 
Using  Eqs. (\ref{eq3})  and (\ref{a2}) we obtain the following equation for $\phi$ in terms of the number of e-folds $\ln a=N$ (setting $M_P=1$ for simplicity)
\begin{equation}\label{AE1}
\phi \frac{d\phi}{d N}=
-\frac{V_{,\phi}}{V}
+\frac{6 q_{0}^{2} c}{V_{,\phi} } e^{-4c \int (V/V_{,\phi}) d\phi}
e^{-4 N} \, .
\end{equation}
Now, it is suitable to rearrange Eq. (\ref{AE1}) to the following form
\begin{equation}
4 c e^{4 N} e^{4 c \int (V/V_{,\phi}) d\phi} 
\Big(1+\frac{V}{V_{,\phi}}\frac{d\phi}{d N}\Big)= 
24 c^2 {q_{0}}^2 \Big(\frac{V}{V_{,\phi}^2}\Big) \, .
\end{equation}
Defining $\mathcal{G}(N)\equiv e^{4 N} e^{4 c \int (V/V_{,\phi}) d\phi}$, the above equation 
takes the following form
\begin{equation}
\frac{d\mathcal{G}}{d N}+4(c-1)\mathcal{G}
= 24 c^2 {q_{0}}^2 \Big(\frac{V}{V_{,\phi}^2}\Big) \,.
\end{equation}
One can solve this differential equation in slow-roll limit to obtain
\begin{equation}\label{as1}
\mathcal{G}(N) = \frac{6 c^2 q_{0}^2\mathcal{C}}{(c-1)}
\Big(\frac{V}{V_{,\phi}^2}\Big)
\Big[1+\frac{6 c^2 q_{0}^2\mathcal{C}}{(c-1)}
\Big(\frac{V}{V_{,\phi}^2}\Big) e^{4N (1-c)}\Big] \, ,
\end{equation}
where $\mathcal{C}$ is a constant of integration. We see that for sufficiently small values
of $q_0^2\mathcal C$, the last term in the above bracket falls off during inflation and  
Eq. (\ref{as1}) implies
\begin{equation}\label{eqq2}
\mathcal{G}(N)^{-1} = e^{-4 N} e^{-4c \int \frac{V}{V_{,\phi}} d \phi}
= \frac{(c-1)}{6 c^2 q_{0}^2} \bigg(\frac{V_{,\phi}^2}{V} \bigg) \, .
\end{equation}
Consequently,  $\rho_{A}$ becomes nearly constant during the second phase of inflation, and after straightforward calculations, we obtain
\begin{equation}
R=\frac{ c-1}{4 c^2 } \Big(\frac{V_{,\phi}}{V}\Big)^2 \, .
\end{equation}
Substituting Eq. (\ref{eqq2}) into the modified slow-roll equation (\ref{a2}), we obtain
\begin{equation}\label{Eq20}
3 H \dot{\phi} \approx -\frac{V_{,\phi}}{c} \, .
\end{equation}
This shows that during the second phase of inflation the effective mass squared of the inflaton field $m^2$  is reduced by the  factor $1/ c$ compared to  the first stage  of inflation \cite{Watanabe:2009ct}. 

\begin{figure}
		\includegraphics[width=0.5\linewidth, height=7.5cm]{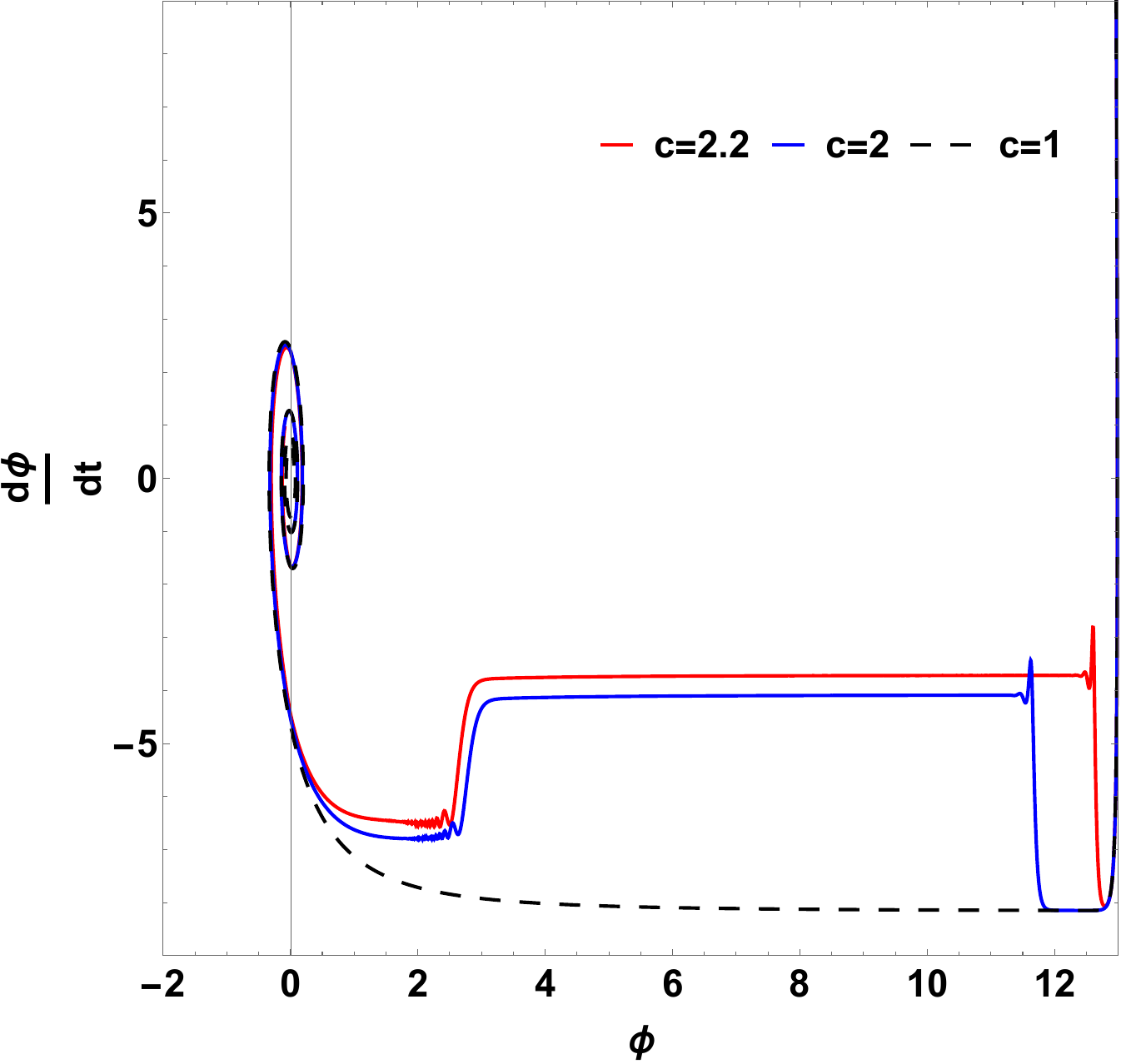} \hspace{0.5cm}
		\includegraphics[scale=0.35]{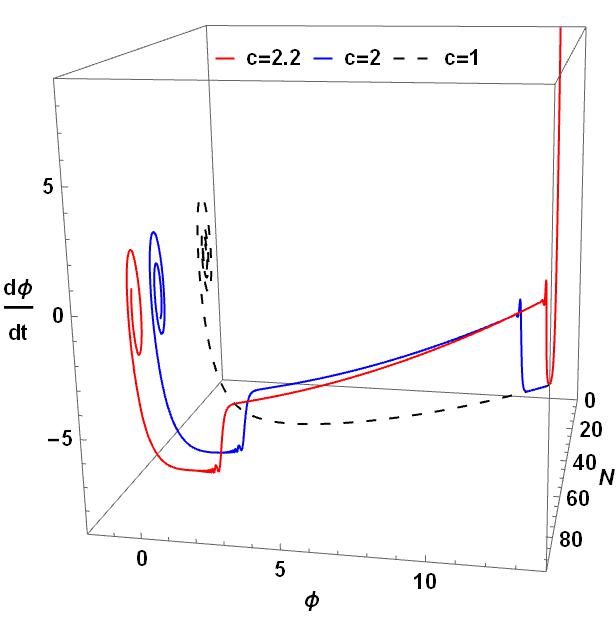} \vspace{0cm}
\caption{Left: The phase space plot of $(\phi, \dot{\phi})$ for the potential $V=\frac{1}{2} m^2 \phi^2$ with parameters $m=10^{-6} M_P$, $\phi(0)=12 M_P$, and $\dot{\phi}(0)=0$. We have fixed $\e=0.01$ and varied the parameter $c$ with three values $c=1, 2$ and $2.2$. The latter two values of $c$ are too large to generate scale invariant power spectrum, but we have chosen them for better visualizations of the effects of gauge fields on inflation dynamics.  
 	Right: The three dimensional plot of $(\phi, \dot{\phi})$ with respect to $N$  for the same parameters as in left figure.  } \label{fig-efixed}
	\vspace{1cm}
	
		\includegraphics[width=0.5\linewidth, height=7.5cm]{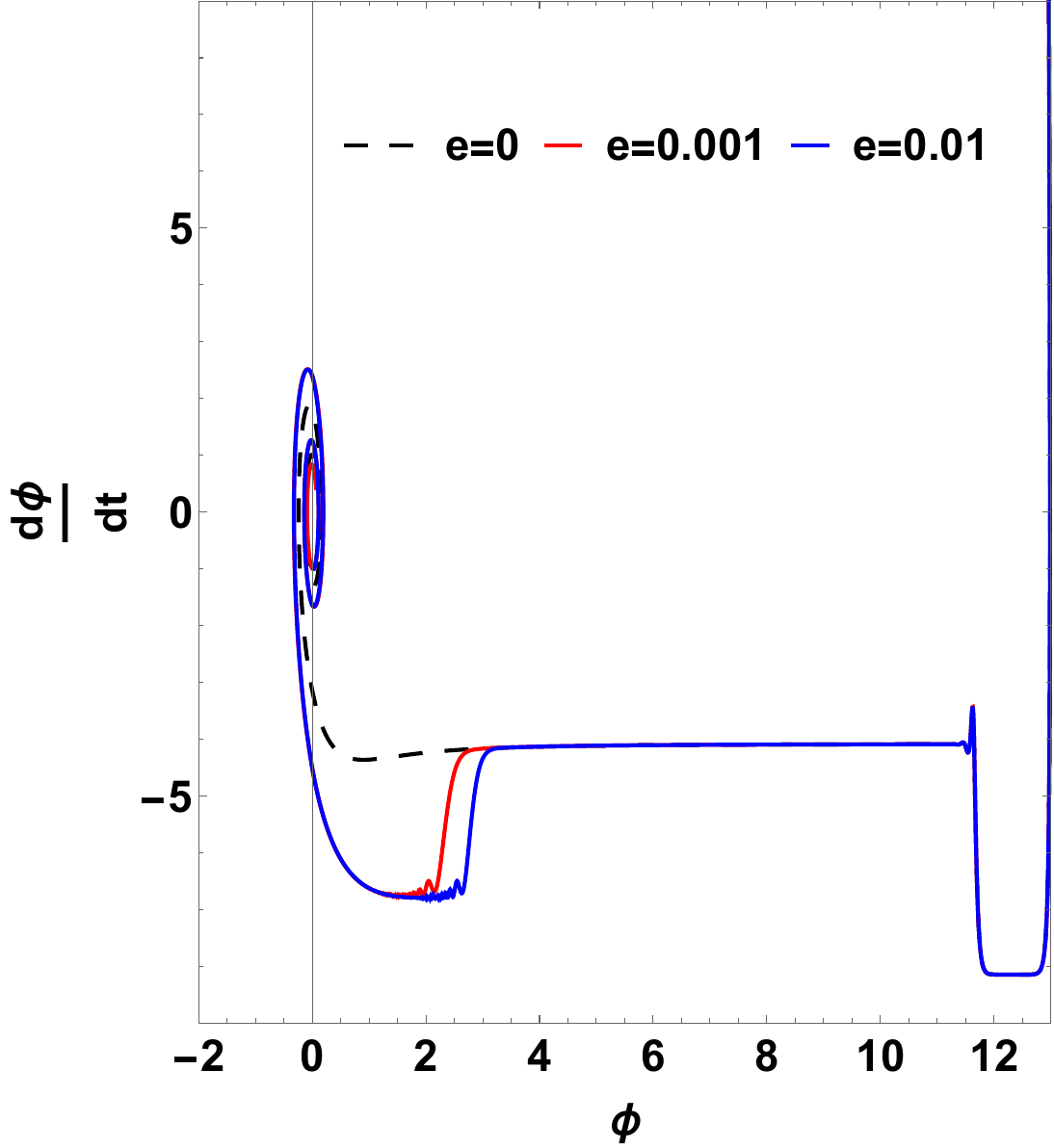}  \hspace{0.5cm}
		\includegraphics[scale=0.35]{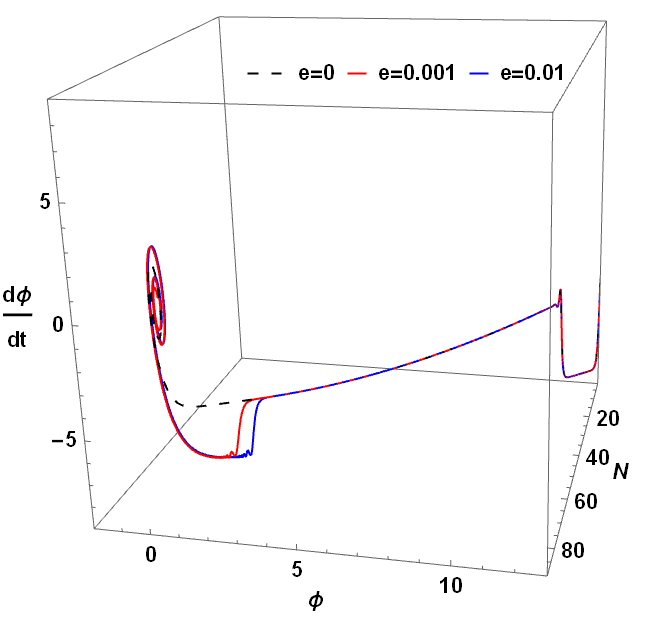}
	\caption{Left: The phase space plot of $(\phi, \dot{\phi})$ with $c$ held fixed at $c=2$ while varying 
	$\e$ with $\e=0, 0.001$ and $0.01$. Other parameters are the same as in top figures. Right: The three dimensional plot of $(\phi, \dot{\phi})$ with respect to $N$  for the same parameters as in left figure. }
\label{fig-cfixed}
\end{figure}

Moreover, from Eqs. (\ref{q1}), (\ref{qq1}) and (\ref{eqq2}), one can also obtain the slow-roll parameter as follows
\begin{equation}\label{Eq21}
\epsilon\equiv -\frac{\dot{H}}{H^2}=\frac{1}{2 c} \Big(\frac{V_{,\phi}}{V}\Big)^2 \,.
\end{equation}
Therefore, we find
\begin{equation}
\label{R2}
{R}=\frac{c-1}{2 c} \epsilon= \frac{I}{2} \epsilon \, ,
\end{equation}
in which we have defined the parameter $I\equiv (c-1)/c$. Interestingly the relation between $R$ and $\epsilon$ given in Eq. (\ref{R2}) is the same as in anisotropic inflation.

In the left panel of Fig. \ref{fig-efixed}, the phase space plot  of  $(\phi ,\dot{\phi})$ for the potential $V=\frac{1}{2} m^2 \phi^2$  for a fixed value of $\e$ and for three different values of $c$ are plotted.
In the right panel of Fig.\ref{fig-efixed} the behaviour of $(\phi ,\dot{\phi})$ as a function of the number of e-folds $N$ is plotted.  As we see from the plots, initially the inflaton field evolves independent of the effects of the gauge field so all three curves coincide during the first phase of inflation. However, as the gauge fields drag enough energy from the background, they kick in and  after a short transient period, the system reaches the attractor phase.  The attractor phase starts sooner for the larger value of $c$. This is understandable, since the larger is the value of $c$, the more energy is pumped into the gauge field from the inflaton field.  We also see that the 
attractor phase and the total number of e-folds are longer for the larger values of $c$. This can be seen from our equations too. Starting from 
$ N=-\int \frac{H}{\dot\phi}d\phi$,  and using Eq. (\ref{Eq20}), we obtain $N=-c\int\frac{V}{V_{,\phi}}d\phi$.   Therefore, the total number of e-fold increases by increasing the value of $c$.

In Fig. \ref{fig-cfixed} the phase space plot  of  $(\phi ,\dot{\phi})$ (left panel) and their dependence on $N$ (right panel) are plotted for the same potential as in Fig. \ref{fig-efixed}, but this time $c$ is held fixed while $\e$ is varied. As can be seen from the plots, $\e$ does not play important roles during much of the period of inflation. However, its effect become important during the final stage of inflation, modifying the total number of e-folds slightly. More specifically, the coupling $\e$ induces an effective mass $m^2\sim\e^2 A^2 e^{-2N}$ for the inflaton field. When this induced mass becomes comparable to  $H$, then the slow-roll conditions are violated and inflation ends abruptly.  During the attractor phase $A\propto e^{(4c-1)N}$ 
so the induced mass scales like $ \e^2 e^{(8c-4) N}$. Consequently, the total number of e-folds depends only logarithmically on $\e$. In other words, holding other parameters such as $c$ fixed while varying $\e$, as in Fig. \ref{fig-cfixed}, the total number of e-folds changes as 
\ba
\Delta N \sim -\frac{1}{2\left(2c-1\right)}\ln \e \, .
\ea
Although $\e$ does not play important roles during the inflation background, but it has important effects on curvature perturbations power spectra and other cosmological observables.

\section{Cosmological Perturbations}
\label{perturbations}

In this section, we present the perturbations of our model based on action 
(\ref{action}). From now on, we work with the conformal time $\tau$ defined as $d \tau = dt/a(t)$.

The metric perturbations around the background geometry (\ref{FRW}) are given by  
\ba\label{metric-perturbations}
\delta{g_{00}} = 2 a^2 \alpha \,, \hspace{.5cm} \delta{g_{0i}} = a^2 
(\partial_i \beta + B_i ) \,,
\hspace{.5cm} \delta{g_{ij}} = a^2 ( 2\psi \delta_{ij} + 2\partial_i\partial_j E 
+ \partial_i F_j + \partial_j F_i +h_{ij} ) \,,
\ea
where $\alpha, \beta, \psi$ and $E$ are scalar modes, $B_i$ and $F_i$ are vector modes while 
$h_{ij}$ are the tensor perturbations which satisfy the following transverse and traceless 
conditions
\ba
\partial_i B_i= \partial_i F_i = \partial_i h_{i j} = h_{ii}=0  \, .
\ea

The gauge fields enjoy internal $O(3)$ symmetry and the perturbations should be defined in 
the spirit of $O(3)$ symmetry as \cite{Emami:2016ldl}
\ba\label{GF-perturbations}
\delta{A_0^{(a)}} = Y_a+\partial_a Y \,, \hspace{.5cm}
\delta{A_i^{(a)}} = \delta{Q}\, \delta_{ia}+\partial_i (\partial_a M+M_a)
+ \epsilon_{iab}(\partial_b{U}+U_b) + t_{ia} \,,
\ea
where $(Y,\delta{Q},M,U)$ are scalar modes, $(Y_a,M_a,U_a)$  are vector modes, and
$(t_{ia})$ label the tensor modes associated with the gauge field perturbations which are
subject to the transverse and traceless conditions
\ba\label{transverse-traceless}
 \partial_i Y_i = \partial_i M_i = \partial_i U_i = \partial_i t_{ij} = t_{ii}= 0 \, . 
\ea
In addition to the above perturbations, we also have the inflaton perturbations $ \delta \phi$. 

The gauge freedom associated with the four-dimensional diffeomorphism invariance fixes 
two scalar modes and two vector modes of metric perturbations.  For the scalar modes, we work in the spatially flat 
gauge in which 
\be\label{SFG}
\psi = 0 \,, \hspace{1cm} E = 0 \,,
\ee
while for the vector perturbations we fix the gauge by setting $F_i=0$. 

Apart from the diffeomorphism invariance, the gauge fields enjoy the $U(1)_a$ gauge invariance given by Eq. (\ref{GF-gaugeF}). But, we have already fixed the $U(1)_a$ gauge in choosing the scalar fields to be real, i.e. going to the unitary gauge, yielding to the action (\ref{action}).

In summary, after fixing the gauges associated with the diffeomorphism invariance and local $U(1)_a$ invariance, we have seven scalar degrees of freedom $(\alpha, \beta, \delta \phi, \delta Q, Y, U,M)$, eight vector degrees of freedom $(B_i, U_a, Y_a, M_a) $ and four tensor perturbations $(h_{ij}, t_{ij})$. In total we have 19 physical degrees of freedom. 

Since the model with the action (\ref{action}) enjoys $O(3)$ symmetry,  the scalar, vector 
and tensor perturbations decouples at the linear order of perturbations. Moreover, since our 
setup is isotropic, the vector perturbations decay as usual in an expanding Universe and we 
will not consider them from now on.

\section{Scalar Perturbations}
\label{sec-scalar}

Working in spatially flat gauge (\ref{SFG}) and fixing local gauge symmetry 
(\ref{GF-gaugeF}), we deal with seven scalar modes 
$(\alpha, \beta, Y, \delta{Q}, U, \delta\phi, M)$.  Direct calculations shows that 
$\alpha, \beta$ appear with no time derivatives in the quadratic action and therefore they can 
be substituted from their algebraic equations of motion. Moreover, the contribution coming 
from these non-dynamical modes are slow-roll suppressed \cite{Emami:2013bk, Chen:2014eua} and we 
therefore neglect them. 

The quadratic action for the remaining modes $( Y, \delta{Q}, U, \delta\phi, M)$   is presented in Appendix \ref{App-action}. As discussed there, the contributions of the perturbations $Y$ and $M$ are suppressed during much of the period of inflation and therefore can be neglected. Therefore, the quadratic action for the 
remaining light scalar perturbations in Fourier space is given by 
\begin{eqnarray}
\label{action-canonical}
\nonumber S^{(2)} &=& \frac{1}{2}\int  d\tau d^3 k \Bigg \{ \delta Q_{c}'^2-\Big(k^2-\frac{2}{\tau^2} \Big) \delta Q_{c}^2 +\delta \phi_{c}'^2-\Big[k^2-\frac{1}{\tau^2}\Big(2+4 I\Big) \Big] \delta \phi_{c}^2\\
 &+&U_{c}'^2-\Big(k^2-\frac{2}{\tau^2} \Big) U_{c}^2
+8 \frac{\sqrt{I}}{\tau^2}\Big[2-\frac{\e^2}{9 H^2} \Big(\frac{\tau_{e}}{\tau}\Big)^4\Big] 
\delta Q_{c} \delta \phi_{c}-\frac{8 \sqrt{I}}{\tau} \delta Q_{c}' \delta \phi_{c} \Bigg \} \, ,
\end{eqnarray}
in which a prime indicates the derivative with respect to the conformal time,  $\tau_e$ is the time of end of inflation and  we have defined the canonically normalized fields
\begin{equation}
\label{canonical fields}
\delta Q_c\equiv\sqrt{2}f\delta Q \,, \;\;\;\;\; 
U_c \equiv kfU \,, \;\;\;\;\; \delta \phi_c \equiv a\delta\phi \,.
\end{equation}
We have ignored pure slow-roll corrections i.e. terms containing the slow-roll parameters $\epsilon$ and  
its derivative without the factor I  since they are the same as those coming from the gravitational back-reactions and can be absorbed  into the power spectrum in the absence of  gauge fields. 
In addition, as we shall show later on, $I \ll1$ so we have kept the leading terms of $I$ in the action (\ref{action-canonical}) which turns out to be proportional to $\sqrt I$.      

Form the action (\ref{action-canonical}), we see that the field $U$ is decoupled from the 
other fields. In addition, it did not exist at the background level. Therefore, the field $U$ 
is a pure isocurvature mode. This is unlike the mode $\delta Q$ which is the perturbations 
associated with the diagonal component of $A_i^{(a)}$ which also had a background 
component, given in Eq. (\ref{3-A}). We see that both the scalar field and the diagonal 
component  of $A_i^{(a)}$ contributes to the background energy and interact with each 
other. In this view, we are dealing with a multiple field model of inflation which is studied 
vastly in the literature. In particular, similar to the logic of \cite{Gordon:2000hv}, we 
expect that a combination of the fields $(\delta \phi, \delta Q)$ to play the roles of the 
adiabatic mode while a different combination to play the role of the entropy perturbations. 

\subsection{Adiabatic and entropy decompositions}

In order to find the adiabatic and entropy modes, we first find the comoving curvature 
perturbations $\calR$ from the standard definition
\begin{equation}\label{curvature-perturbation0}
{\cal R} = \psi + H \delta u \,,
\end{equation}
where $\psi$ measures the spatial curvature and $\delta u$ is the velocity potential which is 
defined as $\delta T^{t}_{i} =  (\rho+p)\partial_{i} \delta u$.  Calculating the 
energy-momentum tensor at the linear order of perturbations, and noting that we work in 
spatially flat gauge Eq. (\ref{SFG}),  the comoving  curvature perturbation takes the following form

\ba\label{curvature-perturbation01}
{\cal R}& =& -a H \frac{\sqrt{2} f A' {\delta Q_c} 
+ a\phi' {\delta \phi_c}+(\e^2/9) a^2 A \phi^2 Y}{2 f^2 A'^2+a^2 \phi'^2} .
\ea
We need to substitute the non-dynamical perturbation $Y$ in the above relation from Eq. (\ref{y}).  As discussed in Appendix \ref{App-action},  the contribution of $Y$ in curvature perturbation is subleading during the inflationary stage. Therefore, to leading order, the curvature perturbation takes the following simple form 
\begin{eqnarray}\label{curvature-pertubation1}
\mathcal{R} & = & - \frac{H}{ {\phi'}} 
\Big[
(1-I) {\delta \phi_c} - \sqrt{I} {\delta Q_c}
\Big] \,.
\end{eqnarray}
The above formula is interesting showing that the contribution of each field into the total curvature perturbation is weighted by the fraction of the corresponding field into the total energy density 
\cite{Wands:2000dp, Lyth:2009zz}. Since $I \ll 1$, the dominant contribution into curvature perturbations is given by the inflaton field perturbations $\delta \phi$. But we expect to have subleading contributions from 
the diagonal component of $A_i^{(a)}$ which is given by the fraction $\sqrt I$ in the above formula.

Following the logic of \cite{Gordon:2000hv}, the scalar modes ${\delta \phi_c}$ and 
${\delta Q_c}$ can be decomposed into the adiabatic and entropy components as follows
\ba
\label{adiabtic}
{\delta \sigma_c} &=& \cos\theta {\delta \phi_c} 
+ \sin\theta {\delta Q_c} \,, \\
\label{entropic}
{\delta s_c} &=& - \sin\theta {\delta \phi_c} + \cos\theta {\delta Q_c} \,,
\ea
where we have defined 
\begin{eqnarray}\label{cos-sin}
\cos\theta \equiv  \sqrt{1-I} \,, \hspace{1cm} 
\sin\theta \equiv - \sqrt{I} \,.
\end{eqnarray} 
The canonical  variables $\delta \sigma_c$ and $\delta s_\sigma$  are related to the standard adiabatic and entropy perturbations defined in  \cite{Gordon:2000hv} via
\ba
 \delta\sigma_c = a\,  \delta \sigma \, \quad \quad   \delta{s_c} = a \, \delta s \, .
\ea

Using the decomposition Eq. (\ref{adiabtic}) into Eq. (\ref{curvature-pertubation1}), the comoving curvature 
perturbations is given by
\begin{eqnarray}
\label{curvature-perturbation}
\mathcal{R} & = & - \frac{{H}}{ \dot \phi} \cos\theta\, {\delta \sigma} \,.
\end{eqnarray} 
In the limit $I\rightarrow0$, we have $\cos\theta=1$ and 
Eq. (\ref{adiabtic}) gives ${\delta \sigma} = {\delta \phi}$ in which we  find the well-known result 
$\mathcal{R} =  - \frac{H}{\dot\phi} \delta\phi $ for the 
curvature perturbations. 

Correspondingly, we define the associated normalized entropy perturbation via
\begin{eqnarray}\label{entropy-perturbation}
\mathcal{S}&\equiv&-\frac{{H}}{\dot \phi} \cos\theta \, {\delta s} \,.
\end{eqnarray}

Our final aim is to find the power spectrum for the observable quantities $\mathcal{R}$ 
and $\mathcal{S}$. For this purpose, we rewrite the quadratic action (\ref{action-canonical}) 
in terms of the adiabatic and entropy modes, yielding 
\begin{eqnarray}\label{action-decomposition}
S^{(2)}= \frac{1}{2}\int d\tau d^3 k &\bigg\{&
U_{c}'^2-\Big(k^2-\frac{2}{\tau^2} \Big) U_{c}^2 \nonumber  \\ 
&+& \delta s_{c}^2-\Big[k^2-\frac{2}{\tau^2}\Big(1+ 6I -\frac{4\e^2 I}{9 H^2} 
\Big(\frac{\tau_{e}}{\tau}\Big)^4\Big)\Big] \delta s_{c}^2 \nonumber \\ 
&+&\delta \sigma_{c}'^2-\Big[k^2-\frac{2}{\tau^2}\Big(1- 4I +\frac{4\e^2 I}{9 H^2} 
\Big(\frac{\tau_{e}}{\tau}\Big)^4\Big)\Big] \delta \sigma_{c}^2
\nonumber  \\  
&+&\frac{8\sqrt{I}}{\tau^2}\Big[2-\frac{\e^2}{9 H^2} 
\Big(\frac{\tau_{e}}{\tau}\Big)^{4} \Big]\delta s_{c} \delta \sigma_{c}
- \frac{8 \sqrt{I}}{\tau} \delta s_{c}' \delta \sigma_{c} \bigg\} \, .
\end{eqnarray}
We see that the adiabatic and entropy modes are coupled to each other with the couplings 
proportional to $\sqrt I$. 

We calculate the power spectra of $\calP_\calR$ and  $\calP_{\cal S}$ and their cross-correlation $\calP_{\calR{\cal S} }$ in next subsections.   However, before that, let us consider the perturbation $U$ which is a pure isocurvature mode and does not couple to other modes.  Decomposing $U$ into the creation and the annihilation operators with the Minkowski (Bunch-Davies) initial condition, we have 
\begin{eqnarray}
\nonumber U_c({\textbf{k}})= {{u}}({ {k}}) a_{ \textbf{k}}+{{u^*}}({ {k}})  a_{ \textbf{-k}}^{\dagger} \,\,\,\ ;\,\,\,\ 
{{u}}(k)= \frac{i e^{-i k \tau}}{\sqrt{2 k^3}\tau} \Big(1+i k \tau\Big) \, .
\end{eqnarray}
Correspondingly,  the dimensionless power spectrum for $U=U_c/a$, defined as usual via 
 $\langle U^\dagger(\tau,{\bf k}) U(\tau,{\bf k}') \rangle \equiv
\frac{2\pi^2}{k^3} {\cal P}_{U} (2 \pi)^3\delta^{(3)}({\bf k}-{\bf k}')$, on super-horizon scales  is given by  
\begin{eqnarray}\label{CF-U2}
{\cal P}_{U} = \Big(\frac{H}{2\pi}\Big)^2 \,.
\end{eqnarray}
The above result shows that the scalar mode $U$ behaves like an spectator field with the amplitude $H/2\pi$.

\subsection{Curvature perturbations power spectrum}
\label{curvature-power}

In this subsection we calculate the curvature perturbation power spectrum $\calP_\calR$. 
From Eq. (\ref{curvature-perturbation}) the power spectrum of curvature perturbation at the end of inflation 
$\tau_e$ is given by 
\begin{eqnarray}
\label{CF-R2}
\langle {\cal R}^\dagger(\tau_e,{\bf k})\, {\cal R}(\tau_e,{\bf k}') \rangle =
\left(\frac{H}{\dot{\phi}}\right)^2 
\cos^2\theta \,{\langle {\delta \sigma}^\dagger 
{\delta \sigma} \rangle}\, 
\equiv \frac{2 \pi^2}{k^{3}}  {\cal P}_{\cal R} \, (2 \pi)^3\delta^{(3)}({\bf k}-{\bf k}') \,.
\end{eqnarray}
The leading contribution to curvature perturbation power spectrum comes from the adiabatic mode $\delta \sigma$. However,  the adiabatic and the entropy modes are coupled to each other with the interactions given by the last two terms in the action (\ref{action-decomposition}). Therefore, we also have to calculate the corrections from the entropy mode in $\calP_\calR$. Since we assume $I \ll 1$, this analysis can be done perturbatively using the standard in-in formalism \cite{Weinberg:2005vy}.

The two-point function for the adiabatic  mode is then given by 
\begin{eqnarray}
\label{IN-IN}
\big \langle {\delta \sigma^2 (\tau_e)}  \big \rangle &=&\Big \langle 0 \Big |\Big[\bar{T} 
\exp\Big(i \int_{\tau_{0}}^{\tau_{e}} H_{I}(\tau'') d \tau'' \Big)\Big] \,  {\delta \sigma}(\tau_e)^2  \, 
\Big[T \exp\Big(-i \int_{\tau_{0}}^{\tau_{e}} H_{I}(\tau') d \tau'\Big)\Big] \Big | 0 \Big \rangle \nonumber\\
&=& \nonumber \langle 0| {\delta \sigma}^2 |0\rangle + i
\Big \langle0 \Big |  \int_{\tau_{0}}^{\tau_{e}} d \tau_{1}\Big[H_{I} (\tau_{1}),\delta \sigma^{2}(\tau_e)\Big] 
  \Big | 0 \Big \rangle
\nonumber\\
&-& \Big \langle0 \Big | \int_{\tau_{0}}^{\tau_{e}} d\tau_{1} \int_{\tau_{0}}^{\tau_{1}} d\tau_{2} \Big[ H_{I} (\tau_{2}),
\Big[ H_{I}(\tau_{1}),{\delta \sigma}^2(\tau_e)\Big]\Big]  \Big | 0 \Big \rangle + ...\, ,
\end{eqnarray}
where $\bar{T}$ and $T$ are the time ordered and anti time ordered operators and $H_{I}$ is the 
interaction Hamiltonian. The integrals are taken from the
initial time $\tau_0 \to - \infty$ when the modes are deep inside the horizon to the end of 
inflation $\tau_e \to 0$. The first term in the second line of Eq. (\ref{IN-IN}) is the two-point
function of the adiabatic mode in the absence of interaction determined by the the free action of 
$\delta \sigma$ in  Eq. (\ref{action-decomposition}). This gives the leading contribution to the curvature perturbations power spectrum, denote by $\calP_\calR^{(0)}$, which is given
by
\begin{eqnarray}
\calP_\calR^{(0)} = \frac{H^2}{8\pi^2 \epsilon M_P^2 } \, .
\end{eqnarray}

In obtaining the above result, we have substituted $\langle 0| {\delta \sigma}^2 |0\rangle=
H^2/2k^3$ and $\big(\frac{H}{\dot{\phi}}\big)^2 \cos^2\theta = {1}/{2\epsilon}$. To be more precise, from Eqs. (\ref{Eq20}) and
(\ref{Eq21}) we find $\big(\frac{H}{\dot{\phi}}\big)^2 \simeq  (1+I)/2 \epsilon$. 
On the other hand, from Eq. (\ref{cos-sin}), we find that $\cos^2\theta = 1- I$ and 
therefore  $\big(\frac{H}{\dot{\phi}}\big)^2 \cos^2\theta = {1}/{(2\epsilon)} + {\cal O}(I^2)$. 

To calculate the corrections in curvature perturbations power spectrum we need to obtain the interaction Hamiltonians. In addition to the  two interactions which directly couple the fields $\delta \sigma$ and $\delta s$ (the last line in   action (\ref{action-decomposition}) containing $\sqrt I$ ) we also have  new interactions in the action from  the second and third lines of Eq. (\ref{action-decomposition}) containing $I$. Note that we treat $I$ as the parameter of the perturbations so any term containing this parameter should be treated as interaction compared to the free theory. In total, we have seven interaction Hamiltonians for the scalar perturbations, $ H^s_{I} = \sum_i^7  H^s_{i}$ with
\begin{eqnarray}\label{HI}
\nonumber &&H^s_1 = - \frac{8 \sqrt{I}}{\tau^2} {\delta \sigma_c} {\delta s_c}, 
\hspace{0.5 cm}
H^s_2 = \frac{4 \sqrt{I}}{\tau} {\delta \sigma_c} {\delta s'_c}, \hspace{0.5 cm} 
H^s_3 = \frac{4 \e^2 \sqrt{I}}{9 H^2} \Big(\frac{\tau_{e}^4}{\tau^6}\Big) 
\delta \sigma_c\delta s_c, \hspace{0.5 cm} 
H^s_4 = \frac{12 I}{\tau^2} {\delta \sigma_c^2}\,,  \\ 
&& H^s_5=- \frac{6 I}{\tau^2} {\delta s_c^2}, \hspace{1cm} 
H^s_{6}= -\frac{4 \e^2 I}{9 H^2} \Big(\frac{\tau_{e}^4}{\tau^6}\Big){\delta \sigma_c^2}, 
\hspace{1cm} H^s_{7}= \frac{4 \e^2 I}{9 H^2} \Big(\frac{\tau_{e}^4}{\tau^6}\Big){\delta s_c^2}.
\end{eqnarray}
Note that because of the kinetic coupling $\delta \sigma \delta s'$, the interaction Hamiltonian is not simply 
$-L_I$. One has to calculate the conjugate momenta $p_j$ corresponding to each field $\delta q_j= \{ \delta \sigma,  \delta s \}$ and then construct 
the Hamiltonian using the standard formula $H= \sum_{i} p_j \delta q_j' - L$. Doing this we find that the 
interactions containing $\delta \sigma^2$ and $\delta s^2$ receive additional contributions compared to what one may naively construct using $H_I= -L_I$.  

Let us denote the correction induced from the interactions to the  adiabatic mode correlation  by $\Delta \langle{\delta{\sigma^2}}\rangle$. Looking at Eq. (\ref{IN-IN}), there are two possible ways for the interaction Hamiltonians to contribute in  $\Delta \langle{\delta{\sigma^2}}\rangle$. If the contribution comes from the single Hamiltonian from the second line of Eq. (\ref{IN-IN}), we denote it by 
$\Delta^{(1)} \langle{\delta{\sigma^2}}\rangle_i$, i.e. it is linear in $H^s_i$.   On the other hand, if the contribution comes from the nested integral containing two Hamiltonians in third line of Eq. (\ref{IN-IN}), then we denote it by $\Delta^{(2)} \langle{\delta{\sigma^2}}\rangle_{i j}$, in which the indices  $i, j$ are for $H^s_i(\tau_1)$ and $H^s_j(\tau_2)$ respectively.   

The free wave function for ${M}_{i\textbf{k}}=\big\{{\delta \sigma_c}(k),{\delta s_c}(k) \big\}$ with the Bunch-Davies initial condition, is given by
\begin{eqnarray}
\label{mode-functions}
{M}_{i \textbf{k}}= {{v}}({ {k}}) a_{i \textbf{k}}+{{v}}({ {k}})^{\star}  a_{i \, \textbf{-k}}^{\dagger} \,  ; \quad 
{{v}}(k)= \frac{i e^{-i k \tau}}{\sqrt{2 k^3}\tau} \Big(1+i k \tau\Big) \, .
\end{eqnarray}

To simplify the notation, let us pull out the factor $(2\pi)^3\delta^{(3)}({\bf k}-{\bf k}')$ and denote the corresponding correlations by $\Delta'$. Then, the leading order corrections in  $\Delta \langle{\delta{\sigma^2}}\rangle$ are obtained to be
\begin{eqnarray}
\label{correction-dsigma-4}
\Delta'^{(1)} \langle{\delta \sigma_c^2}\rangle_4
&=& i \int_{\tau_{0}}^{\tau_{e}} d \tau_{1}\Big[H_{4} (\tau_{1}),\delta \sigma_c^{2}(\tau)\Big]
=-48 I\, 
{\rm Re}\Big[ i  \int_{\tau_{0}}^{\tau_{e}} d \tau_{1} \Big(\frac{1}{\tau_{1}}\Big)^2 \Big({{v}} (\tau_{1}) {{v}}^{\star } (\tau_{e})\Big)^2\Big] \nonumber \\
&=& \frac{8I N_e}{k^3 \tau_e^2} \,  ,
\end{eqnarray}

\begin{eqnarray}\label{correction-dsigma-11}
\nonumber \Delta'^{(2)} \langle{\delta \sigma_c}^2\rangle_{11}
&=&512 I \int_{\tau_{0}}^{\tau_{e}} d \tau_{1}\int_{\tau_{0}}^{\tau_{1}} d \tau_{2} \Big(\frac{1}{\tau_{1} \tau_{2}}\Big)^2 {\rm Im}\Big[{{v}}(\tau_{1}) {{v}}^{\star}(\tau_{e})\Big] {\rm Im} \Big[{{v}}(\tau_{2}) \bar{{v}}^{\star}(\tau_{e}){{v}}(\tau_{2}) {{v}}^{\star}(\tau_{1})\Big]\\
&=&\frac{64 I N_e^2}{9 k^3\tau_e^2}  \,,
\end{eqnarray}

\begin{eqnarray}\label{correction-dsigma-12}
\nonumber \Delta'^{(2)} \langle{\delta \sigma_c}^2\rangle_{12}
&=&-256 I \int_{\tau_{0}}^{\tau_{e}} d \tau_{1}\int_{\tau_{0}}^{\tau_{1}} d \tau_{2} \Big(\frac{1}{\tau_{1}^2 \tau_{2}}\Big) {\rm Im}\Big[v(\tau_{1}) v^{\star}(\tau_{e})\Big] {\rm Im} \Big[v(\tau_{2}) v^{\star}(\tau_{e})v'(\tau_{2}) v^{\star}(\tau_{1})\Big]\\
&=&-\frac{16 I N_e^2}{9 k^3 \tau_e^2} \,,
\end{eqnarray}

\begin{eqnarray}\label{correction-dsigma-21}
\nonumber \Delta'^{(2)} \langle{\delta \sigma_c}^2\rangle_{21}
&=&-256 I \int_{\tau_{0}}^{\tau_{e}} d \tau_{1}\int_{\tau_{0}}^{\tau_{1}} d \tau_{2} \Big(\frac{1}{\tau_{1} \tau_{2}^2}\Big) {\rm Im}\Big[v(\tau_{1}) v^{\star}(\tau_{e})\Big] {\rm Im} \Big[v(\tau_{2}) v^{\star}(\tau_{e})v(\tau_{2}) v'^{\star}(\tau_{1})\Big]\\
&=&\frac{32 I N_e^2}{9 k^3 \tau_e^2}   \,,
\end{eqnarray}

\begin{eqnarray}\label{correction-dsigma-22}
\Delta'^{(2)} \langle{\delta \sigma_c}^2\rangle_{22}
&=&128 I \int_{\tau_{0}}^{\tau_{e}} d \tau_{1}\int_{\tau_{0}}^{\tau_{1}} d \tau_{2} \Big(\frac{1}{\tau_{1} \tau_{2}}\Big) {\rm Im}\Big[v(\tau_{1}) v^{\star}(\tau_{e})\Big] {\rm Im} \Big[v(\tau_{2}) v^{\star}(\tau_{e})v'(\tau_{2}) v'^{\star}(\tau_{1})\Big] \nonumber \\
&=&-\frac{ 8 I N_e^2}{9 k^3 \tau_e^2} \,,
\end{eqnarray}

\ba\label{correction-dsigma-33} 
\Delta'^{(2)} \langle{\delta \sigma_c}^2\rangle_{33}&=&\frac{128 I\e^4}{81 H^4} \int_{\tau_{0}}^{\tau_{e}} d \tau_{1}\int_{\tau_{0}}^{\tau_{1}} d \tau_{2} \Big(\frac{\tau_e^8}{\tau_{1}^6 \tau_{2}^6}\Big) {\rm Im}\Big[{{v}}(\tau_{1}) {{v}}^{\star}(\tau_{e})\Big] {\rm Im} \Big[{{v}}(\tau_{2}) {{v}}^{\star}(\tau_{e}){{v}}(\tau_{2}) {{v}}^{\star}(\tau_{1})\Big]
\nonumber\\
&=&\frac{I \e^4}{4851 H^4 k^3\tau_e^2} \,,
\ea
\ba\label{correction-dsigma-31} 
\Delta'^{(2)} \langle{\delta \sigma_c}^2\rangle_{31}&=&-\frac{256 I \e^2}{9 H^2}  \int_{\tau_{0}}^{\tau_{e}} d \tau_{1}\int_{\tau_{0}}^{\tau_{1}} d \tau_{2} \Big(\frac{\tau_e^4}{\tau_{1}^6\tau_{2}^2}\Big) {\rm Im}\Big[{{v}}(\tau_{1}) {{v}}^{\star}(\tau_{e})\Big] {\rm Im} \Big[{{v}}(\tau_{2}) {{v}}^{\star}(\tau_{e}){{v}}(\tau_{2}) {{v}}^{\star}(\tau_{1})\Big]
\nonumber\\
&=&-\frac{16 I \e^2 N_{e}}{189  k^3H^2\tau_e^2} \,,
\ea
\ba\label{correction-dsigma-32} 
\Delta'^{(2)} \langle{\delta \sigma_c}^2\rangle_{32}&=&\frac{128 I \e^2}{9 H^2}  \int_{\tau_{0}}^{\tau_{e}} d \tau_{1}\int_{\tau_{0}}^{\tau_{1}} d \tau_{2} \Big(\frac{\tau_e^4}{\tau_{1}^6 \tau_{2}}\Big) {\rm Im}\Big[{{v}}(\tau_{1}) {{v}}^{\star}(\tau_{e})\Big] {\rm Im} \Big[v(\tau_{2}) v^{\star}(\tau_{e})v'(\tau_{2}) v^{\star}(\tau_{1})\Big]
\nonumber\\
&=&\frac{4I \e^2 N_{e}}{189 k^3 H^2\tau_e^2} \,,
\ea
where $N_{e} = -\ln(-k \tau_{e})$ is the number of e-folds at the end of inflation and $\Delta'^{(1)} \langle{\delta \sigma_c}^2\rangle_{5}= \Delta'^{(2)} \langle{\delta \sigma_c}^2\rangle_{13}=\Delta'^{(2)} \langle{\delta \sigma_c}^2\rangle_{23}=0$. Note that with $N_e \sim 50-60$  we have neglected the sub-leading corrections containing $I N_e$ compared to $I N_e^2$ in the last 
 nested integrals above. 

 Now, combining the above results, and neglecting the subleading $I N_e$ contributions against the $I  N_e^2$ contributions, the total curvature perturbation power spectrum is obtained to be 
\ba
\label{PS-R}
\mathcal{P}_{\mathcal{R} } = {\calP}^{(0)}_{\mathcal{R}}\left(1+16IN_e^2F(\beta)\right) \,,
\ea
with 
\ba\label{betA}
\beta \equiv  \frac{\e^2 M_P^2}{126 H^2N_e},\,\,\,\,\,\,\,\,\, F(\beta) \equiv1- \beta+\frac{9}{22}\beta^2 \,.
\ea
The parameter $\beta$ measures the effects of the gauge coupling $\e^2$. With $M_P/H \sim 10^{5}$, we have $\beta \gtrsim 1$ for $\e \gtrsim 10^{-3}$. For large value of $\e$ the function $F(\beta)$ grows like $\beta^2$. 

Interestingly, the correction from the gauge field dynamics in curvature perturbations in 
Eq. (\ref{PS-R})   has the same form as in \cite{Chen:2014eua} studied in the context of charged anisotropic inflation model.   However, in the model of \cite{Chen:2014eua} with a single copy of $U(1)$ gauge field, 
the gauge field corrections in  power spectrum induce statistical anisotropy $\Delta \calP_\calR/{ \calP_\calR}^{(0)}= g_* \cos^2 ({\bf \hat k} \cdot {\bf \hat n})$ with the quadrupolar amplitude $g_* = -24 I F(\beta) N_e^2$
 in which ${\bf \hat n}$ is the preferred direction (direction of anisotropy) in the sky. Note that when $\e=\beta=0$, then $F(\beta)=1$ and one recovers the well known  results \cite{Kanno:2010ab, Bartolo:2012sd, Emami:2013bk, Abolhasani:2013zya}  $g_* = -24 I N_e^2$. In order to be consistent with the observational constraints  $|g_*| \lesssim 10^{-2}$ \cite{Ade:2015hxq, Kim:2013gka}, one then requires $I \lesssim 10^{-7}$. 
However, in our setup with internal $O(3)$ symmetry,  we have three orthogonal gauge fields with equal amplitude  so there is no statistical anisotropy.  As a result, we have less stringent constraint on the value of $I$.

Having calculated the corrections in curvature perturbation power spectrum, we can also calculate the corrections in the spectral index $\Delta  n_s$, given by
\ba
\label{ns}
\Delta n_s  = \Delta \frac{d \ln \calP_\calR}{d \ln k}\Big |_*&=& \Big(32 I N_e F(\beta)+16I\beta\big(-1+\frac{9}{11}\beta\big)\Big)\frac{d N_e}{ d \ln \, k}\nonumber\\
& =& \Big(32 I N_e F(\beta)-16 I N_e\beta\big(-1+\frac{9}{11}\beta\big)\Big)\, ,
\ea 
in which the subscript $*$ represents the time of horizon crossing for the mode of interest $k$. 

In order to have a nearly scale invariant power spectrum we require $\Delta n_s $ to be at the order of the slow-roll parameters. As a result, we conclude that $I \lesssim \epsilon/10 N_e$. This justifies our assumption in taking $I \ll 1$. However, the above result also indicates that $I$ is parametrically at the order $I \sim 10^{-2} \epsilon \sim 10^{-4}$ assuming that $\epsilon$ is at the order of few percent. This is less restrictive compared to constraint imposed on the magnitude of $I$ in models of anisotropic inflation discussed above.

The smallness of $I$ may raise concerns about the existence of the background attractor regime \cite{Naruko:2014bxa, Fujita:2017lfu}.   One may require some fine-tunings on the combination $q_0^2\mathcal C$ in order to neglect the last 
term in the brackets in (\ref{as1}). To be specific, for the chaotic inflation with 
$V=\frac{1}{2}m^2\phi^2$, the condition $\frac{6 c^2 q_{0}^2\mathcal{C}}{(c-1)} 
\Big(\frac{V}{V_{,\phi}^2}\Big) e^{4N (1-c)} \ll 1$ requires

\begin{equation}
q_0^2\mathcal C <  I \epsilon  \sim \epsilon^2/N \sim \epsilon^3.
\end{equation}
This indicates the level of fine-tuning required in order for the gauge field dynamics to actually reach the attractor phase.


\subsection{ $\calP_{\cal S}$ and  $\calP_{\calR{\cal S} }$}
\label{entropy-power-sec}

In this subsection we calculate the power spectrum of entropy mode  $\calP_{\cal S}$ and its 
cross-correlation with the curvature perturbation $\calP_{\calR{\cal S} }$.

For the cross-correlation, we find 
\begin{eqnarray}\label{dsigma-ds-CC}
\Delta'^{(1)} \langle {\delta \sigma_c} {\delta s_c}\rangle 
&=& i \int_{\tau_{0}}^{\tau_{e}} d \tau_{1}
\Big[H_{1}^s (\tau_{1})+H_{2}^s (\tau_{1})+H_{3}^s (\tau_{1}),
{\delta \sigma_c} {\delta s_c} (\tau_e)\Big] \nonumber \\
 &=&16 \sqrt{I}\, {\rm Re}\Big[ i  \int_{\tau_{0}}^{\tau_{e}}d\tau_{1} \Big(\frac{1}{\tau_{1}}\Big)^2 v (\tau_{1})^2 v_{2}^{\star } (\tau_{e})^2 \Big] \nonumber \\
 &-& 8 \sqrt{I}\, {\rm Re}\Big[ i  \int_{\tau_{0}}^{\eta_{e}} d\tau_{1}\Big(\frac{1}{\tau_{1}}\Big) v' (\tau_{1}) v^{\star } (\tau_{e}) 
v(\tau_{1}) v^{\star} (\tau_{e})\Big] \nonumber\\
 &-&8 \sqrt{I}\frac{\e^2\tau_e^4}{9H^2}\, {\rm Re}\Big[ i  \int_{\tau_{0}}^{\tau_{e}}d\tau_{1} \Big(\frac{1}{\tau_{1}}\Big)^6 
v(\tau_{1})^2 v^{\star } (\tau_{e})^2\Big] \nonumber\\
 &=& -\frac{2\sqrt{I} N_e}{k^3 \tau_e^2} + \frac{ \e^2 
\sqrt{I} }{63 H^2 k^3\tau_e^2} \,.
\end{eqnarray}
We see that, unlike in previous integrals, the cross-correlation is proportional to $\sqrt{I} $. The reason is that we did not have to calculate a nested integral.  Correspondingly, the cross-correlation of the entropy and the curvature perturbation is given by 
\begin{eqnarray}
\label{PS-SR}
\mathcal{P}_{ \mathcal{R}\mathcal{S}}
= - 4 \sqrt{I} \calP_\calR^{(0)} N_e (1 - \beta) \,.
\end{eqnarray}

To calculate $\calP_{\cal S}$ we can perform similar in-in integrals as in the case of curvature perturbations in previous subsection. However, there is a less cumbersome way to obtain $\calP_{\cal S}$ as we describe below. Let us first look at the interaction Hamiltonians 
$H^s_1$ and $H^s_2$ which are given by Eq. (\ref{HI}). We can perform an integration by 
part and find $H^s_1+H^s_2 = - \frac{4 \sqrt{I}}{\tau^2} {\delta \sigma_c} {\delta s_c} 
- \frac{4 \sqrt{I}}{\tau} {\delta \sigma'_c} {\delta s_c}$. Now we make the identification 
$\delta\sigma_c \leftrightarrow \delta {s}_c$ with
\ba
H^{s}_1 \leftrightarrow \frac{1}{2} H^{s}_1 \,, \hspace{1cm}
H^{s}_2 \leftrightarrow - H^{s}_2 \,,
\ea
from which we can easily find
\begin{eqnarray}
\label{correction-ds-explicit}
&& \Delta'^{(2)} \langle{\delta s_c}^2\rangle_{11} 
= \frac{1}{4} \times \Delta'^{(2)} \langle {\delta{\sigma_c^2}}\rangle_{11} 
= \frac{16 I N_e^2}{9 k^3\tau_e^2} \,, \\
&& \Delta'^{(2)} \langle{\delta s_c}^2\rangle_{12} 
= - \frac{1}{2} \times \Delta'^{(2)} \langle {\delta{\sigma_c^2}}\rangle_{12} 
= \frac{8 I N_e^2}{9 k^3\tau_e^2} \,, \nonumber \\
&& \Delta'^{(2)} \langle{\delta s_c}^2\rangle_{21} 
= - \frac{1}{2} \times \Delta'^{(2)} \langle{\delta{\sigma_c^2}}\rangle_{21} 
= - \frac{16 I N_e^2}{9 k^3\tau_e^2} \,, \nonumber \\
&& \Delta'^{(2)} \langle{\delta s_c}^2\rangle_{22} 
= \Delta'^{(2)} \langle {\delta{\sigma_c^2}}\rangle_{22} 
= - \frac{8  I N_e^2}{9 k^3\tau_e^2} \,.
\end{eqnarray}

Summing up all the above corrections, we see that they neatly cancel each other and 
therefore we do not have any $I N_e^2$ correction to the power spectrum of the entropy
mode. We have already seen that $H^s_4$ gives corrections at the order $IN_e$ to the 
curvature perturbation power spectrum which we have neglected in comparison with the
$I N_e^2$ corrections. Here, however, we have to consider it since there is no 
$I N_e^2$ correction. The $IN_e$ correction to the entropy mode comes from the
interaction Hamiltonian $H^s_5$. From Eq. (\ref{HI}) we can see that we should consider the
following identification
\ba
H^{s}_5 \leftrightarrow -\frac{1}{2} H^{s}_4 \,,
\ea
which implies
\begin{eqnarray}
\Delta'^{(1)} \langle{\delta s_c}^2\rangle_{5} 
= - \frac{1}{2} \times \Delta'^{(1)} \langle {\delta{\sigma_c^2}}\rangle_{4} 
= - \frac{4 I N_e}{k^3\tau_e^2} \,.
\end{eqnarray}

From Eq. (\ref{HI}), it is clear that the interaction Hamiltonians $H^s_1$ and $H^s_3$ are symmetric in
$\delta \sigma_c \leftrightarrow \delta s_c$. Therefore
we simply have
\begin{eqnarray}
\Delta'^{(2)} \langle{\delta s_c}^2\rangle_{31} 
= \Delta'^{(2)} \langle {\delta{\sigma_c^2}}\rangle_{31} 
= -\frac{16 I \e^2 N_{e}}{189  k^3H^2\tau_e^2} \,, \\ \nonumber
\Delta'^{(2)} \langle{\delta s_c}^2\rangle_{33} 
= \Delta'^{(2)} \langle {\delta{\sigma_c^2}}\rangle_{33} 
= \frac{I \e^4}{4851 H^4 k^3\tau_e^2} \,.
\end{eqnarray}

The last correction to the power spectrum of the entropy mode comes from the interaction 
Hamiltonians $H^s_2$ and $H^s_3$. Performing an integration by part, it is easy to see that
the appropriate identification will be
\ba
H^{s}_2 \leftrightarrow - H^{s}_2 - \frac{1}{2} H^{s}_1 \,, \hspace{1cm}
H^{s}_3 \leftrightarrow H^{s}_3 \,,
\ea  
which gives
\begin{eqnarray}
\Delta'^{(2)} \langle{\delta s_c}^2\rangle_{32} 
= - \Delta'^{(2)} \langle {\delta{\sigma_c^2}}\rangle_{32} 
- \frac{1}{2} \times \Delta'^{(2)} \langle {\delta{\sigma_c^2}}\rangle_{31} 
= \frac{4 I \e^2 N_{e}}{189  k^3 H^2\tau_e^2} \,.
\end{eqnarray}

In the same manner we can easily see $\Delta'^{(2)} \langle{\delta s_c}^2\rangle_{13}=
\Delta'^{(2)} \langle{\delta s_c}^2\rangle_{23}=0$. 
 
All of these results can also be confirmed 
from the direct in-in calculations. Summing up all the above corrections, we find
\begin{eqnarray}
\label{PS}
\mathcal{P}_{ \mathcal{S}} =  \calP_\calR^{(0)}
\Big[ 1 - 8 I N_e + 16 I N_e^2 \Big(F(\beta) -1 \Big) \Big] \,,
\end{eqnarray}
where $\beta$ and $F(\beta)$ are defined in Eq. (\ref{betA}).


\section{Tensor Perturbations}
\label{sec-tensor}

There are two different types of tensor perturbations in our model. One is the usual tensor perturbations of the metric $h_{ij}$. The other one is $t_{ij}$  coming from the matter sector of  the $O(3)$ gauge fields in Eq. (\ref{GF-perturbations}). We therefore have four tensor modes  in our model. 

Using the transverse and traceless conditions, the quadratic action in Fourier space
is obtained to be
\begin{eqnarray}\label{action2-tensor}
S^{(2)}=\frac{1}{2}\int d^3 k d\tau 
&\Bigg\{ &
{\bar{h}_{ij}}'^2 - \Big(k^2 - \frac{2+2I\epsilon}{\tau^2} \Big) \bar{h}_{ij}^2
+ {\bar{t}_{ij}}'^2 - \Big(k^2 - \frac{2-5I\epsilon}{\tau^2} \Big) \bar{t}_{ij}^2
\nonumber \\  
&+& \frac{4 \sqrt{I\epsilon}}{\tau^2} 
\big( \tau \bar{h}_{ij} {\bar{t}_{ij}}' - 2 \bar{h}_{ij} \bar{t}_{ij} \big)
+ \frac{8 \sqrt{I\epsilon}\, \e^2}{9 \tau^2 H^2\epsilon} \Big(\frac{\tau_{e}}{\tau}\Big)^4
\bar t_{ij}\bar h_{ij} \Bigg\} \,,
\end{eqnarray}
where we have defined the canonically normalized fields as follows
\begin{equation}\label{tensor-canonical}
\bar{h}_{ij} \equiv \frac{a}{2} \, h_{ij} \,, \hspace{1cm} 
\bar{t}_{ij} \equiv f \, t_{ij} \,.
\end{equation}

It is convenient to write the tensor modes in terms of their polarizations. In order to do this, we 
note that the traceless and transverse conditions  imply ${\bar h}_{ii}
=k_i {\bar h}_{ij}= {\bar t}_{ii} =k_i {\bar t}_{ij} = 0$. Consequently, we can  express them in terms of the polarization tensor as ${\bar h}_{ij} = 
\sum_{+,\times} \bar{h}^{\lambda} e^{\lambda}_{ij}$ and ${\bar t}_{ij} = \sum_{+,\times} 
\bar{t}^{\lambda} e^{\lambda}_{ij}$ where we have $e^\lambda_{ii} = k^i e^\lambda_{ij} = 0$ and 
$e^\lambda_{ij} e^{\lambda'}_{ij} = 2 \delta_{\lambda\lambda'}$.

The interaction terms in (\ref{action2-tensor}) are proportional to $\sqrt{I\epsilon}$. In the
previous section, we have seen that $I \lesssim 10^{-2} \epsilon$ and therefore 
$\sqrt{I\epsilon} \lesssim  \epsilon/10$ which is small. On the other hand, the interactions in 
(\ref{action2-tensor}) have the same form as the interactions in  (\ref{action-decomposition}). Therefore,  from our results for the scalar modes, the leading corrections in tensor correlations are 
at the order  $I\epsilon N_e^2$. 

The wave functions for the free tensor modes ${N}_{i\textbf{k}} = \{ \bar{h}^{\lambda}(k),\bar{t}^{\lambda}(k) \}$ are given by
\begin{eqnarray}
{N}_{i\textbf{k}}= n({{k}}) a_{i \textbf{k}}+n({{k}})^*  a_{i\,  \textbf{-k}}^{\dagger} \,\,\,\ ; \quad 
n(k)= i\frac{e^{-i k \tau}}{\sqrt{2 k^3}\tau} \Big(1+i k \tau\Big) \,.
\end{eqnarray}

The interaction Hamiltonians associated with the quadratic action (\ref{action2-tensor}) 
in the interaction picture are given by 
\begin{eqnarray}\label{int-H-tensor}
&&H^t_1 = \frac{8 \sqrt{I\epsilon}}{\tau^2} \sum_{+,\times} \bar{h}^{\lambda} 
\bar{t}^{\lambda} \, , \hspace{.5cm}
H^t_2 = -\frac{4 \sqrt{I \epsilon}}{\tau} \sum_{+,\times} 
\bar{h}^{\lambda}{\bar{t}'^{\lambda}}\,, \hspace{.5cm} 
H^t_3 = -\frac{8 \sqrt{I\epsilon}\, \e^2}{9 \tau^2 H^2 \epsilon} 
\Big(\frac{\tau_{e}}{\tau}\Big)^4 \sum_{+,\times} 
\bar{h}^{\lambda} \bar{t}^{\lambda}\,, \nonumber \\
&&H^t_4 = \frac{2I\epsilon}{\tau^2} \sum_{+,\times} 
\bar{h}^{\lambda} \bar{h}^{\lambda}\,, \hspace{.7cm}
H^t_5 = \frac{5I\epsilon}{\tau^2} \sum_{+,\times} 
\bar{t}^{\lambda} \bar{t}^{\lambda}\,.
\end{eqnarray}

Similar to the analysis of entropy power spectrum in subsection \ref{entropy-power-sec},  we do not
need to explicitly perform the cumbersome in-in calculations since we can simply model
the above interaction Hamiltonians to those we had in the case of scalar perturbations
given in Eq. (\ref{HI}) via the following identifications
\begin{eqnarray}\label{int-H-t-Identify}
H^{t}_1 \leftrightarrow -\sqrt{\epsilon} H^{s}_1\,, \hspace{.5cm}
H^{t}_2 \leftrightarrow -\sqrt{\epsilon} H^{s}_2\,, \hspace{.5cm}
H^{t}_3 \leftrightarrow -\frac{2}{\sqrt{\epsilon}} H^{s}_3\,, \hspace{.5cm}
H^{t}_5 \leftrightarrow -\frac{5}{6} \epsilon H^{s}_5\,.
\end{eqnarray}
Using the above identifications and the results obtained from Eq. (\ref{correction-dsigma-11}) to Eq. (\ref{correction-dsigma-32}),  we can easily obtain the nonzero corrections to the power spectrum of the tensor modes as follows 
\begin{eqnarray}
 \nonumber \Delta'^{(2)} \langle ({\bar{h}^{\lambda}})^2\rangle_{11}
 &=& \epsilon \, \Delta'^{(2)} \langle \delta\sigma_c^2\rangle_{11}
= \frac{64 I \epsilon N_e^2}{9 \tau_e^2 k^3}\,, \\
 \nonumber \Delta'^{(2)} \langle ({\bar{h}^{\lambda}})^2\rangle_{12}
&=& \epsilon \, \Delta'^{(2)} \langle \delta\sigma_c^2 \rangle_{12}
=- \frac{16 I \epsilon N_e^2}{9 \tau_e^2 k^3}\, \,, \\
\nonumber \Delta'^{(2)} \langle ({\bar{h}^{\lambda}})^2\rangle_{21}
&=& \epsilon \, \Delta'^{(2)} \langle \delta\sigma_c^2\rangle_{21}
= \frac{32 I \epsilon N_e^2}{9 \tau_e^2 k^3}\,, \\ \nonumber 
\Delta'^{(2)} \langle ({\bar{h}^{\lambda}})^2\rangle_{22}
&=& \epsilon \, \Delta'^{(2)} \langle \delta\sigma_c^2 \rangle_{22}
= - \frac{8  I \epsilon N_e^2}{9 \tau_e^2 k^3}\, \,, \\
\nonumber \Delta'^{(2)} \langle ({\bar{h}^{\lambda}})^2\rangle_{33} 
&=& \frac{4}{\epsilon} \Delta'^{(2)} \langle \delta\sigma_c^2\rangle_{33}
=\frac{4 \e^4 (I/\epsilon)}{4851  H^4 k^3 \tau_{e}^2} \,, \\
\nonumber \Delta'^{(2)} \langle ({\bar{h}^{\lambda}})^2\rangle_{31} 
&=& 2 \Delta'^{(2)} \langle \delta\sigma_c^2\rangle_{31}
=-\frac{32 \e^2 I N_e}{189 H^2 k^3 \tau_{e}^2} \,, \\
\nonumber \Delta'^{(2)} \langle ({\bar{h}^{\lambda}})^2\rangle_{32} 
&=& 2 \Delta'^{(2)} \langle \delta\sigma_c^2\rangle_{32}
= \frac{8 \e^2 I  N_{e}}{189  H^2 k^3 \tau_{e}^2} \,,
\end{eqnarray}
with $\Delta'^{(2)} \langle ({\bar{h}^{\lambda}})^2\rangle_{13}=\Delta'^{(2)} 
\langle ({\bar{h}^{\lambda}})^2\rangle_{23}=0$. 

Summing up all the above corrections we find
\ba\label{hh-correction}
\Delta'^{(2)} \langle \bar{h^{\lambda}}^{\dagger} \bar{h}^{\lambda'}\rangle
&=& \frac{8I}{k^3 \tau_e^2} \Big( \epsilon N_e^2 - \frac{\e^2 N_e}{63 H^2}
+ \frac{(\e^4/\epsilon)}{9702 H^4} \Big) \delta_{\lambda \lambda'} \,.
\ea
Note the important effect that  the charge coupling interaction induces $1/\epsilon$ enhancement  to the tensor power spectrum which is the specific feature of this model. This is similar to the results obtained in  model of charged anisotropic inflation \cite{Chen:2014eua} where the statistical anisotropy induced in tensor power spectrum is more pronounced compared to statistical anisotropy induced in the scalar power spectrum. 

To calculate the power spectrum of the gauge field tensor mode, we note that it appears
exactly the same as entropy mode. Therefore, upon making the appropriate identifications of the interaction Hamiltonians,  we find the following
results
\begin{eqnarray}
\nonumber \Delta'^{(2)} \langle ({\bar{t^{\lambda}}})^2\rangle_{11} 
&=& \frac{64 I \epsilon N_e^2}{9k^3\tau_e^2} \,, \hspace{1cm}
\Delta'^{(2)} \langle ({\bar{t^{\lambda}}})^2\rangle_{12}=-
\frac{16 I \epsilon N_e^2}{9 k^3\tau_e^2} \,, \\ \nonumber 
\Delta'^{(2)} \langle ({\bar{t^{\lambda}}})^2\rangle_{21} 
&=& - \frac{64 I \epsilon N_e^2}{9 k^3\tau_e^2}  \,, \hspace{1cm}
\Delta'^{(2)} \langle ({\bar{t^{\lambda}}})^2\rangle_{22} =
\frac{16 I \epsilon N_e^2}{9 k^3\tau_e^2} \,, \\ \nonumber
\Delta'^{(2)} \langle ({\bar{t^{\lambda}}})^2\rangle_{33}
&=&\frac{4 \e^4(I/\epsilon)}{4851 H^4 k^3\tau_e^2} \,, \hspace{1cm}
\Delta'^{(2)} \langle ({\bar{t^{\lambda}}})^2\rangle_{31}
=-\frac{32I N_{e} \e^2}{189 k^3 H^2\tau_e^2} \,,
\\ \nonumber
\Delta'^{(2)} \langle ({\bar{t^{\lambda}}})^2\rangle_{32}
&=&\frac{8I N_{e} \e^2}{189 k^3 H^2\tau_e^2} \,, \hspace{1cm}
\Delta'^{(1)} \langle ({\bar{t^{\lambda}}})^2\rangle_{5}
=\frac{10I \epsilon N_{e}}{3 k^3 \tau_e^2} \,,
\end{eqnarray}
with $\Delta'^{(2)} \langle ({\bar{t^{\lambda}}})^2\rangle_{13}
=\Delta'^{(2)} \langle ({\bar{t^{\lambda}}})^2\rangle_{23}=0$. 

Summing the above corrections, we see that they  cancel one another and, similar to the case of 
$\calP_{{\cal S}}$,  there is no $I\epsilon N_e^2$ correction to the two-point function of ${\bar t}^\lambda$
and we have to keep the  $I\epsilon N_e$  corrections. 

What remain is the cross-correlation between  ${\bar h}^\lambda$ and  ${\bar t}^\lambda$. Keeping the above identifications in mind,  looking at Eq. (\ref{dsigma-ds-CC}), we see that the first term in the last line comes from  the interaction Hamiltonians $H^s_1$ and $H^s_2$ in Eq. (\ref{HI}) while the second term  comes from $H^s_3$ in Eq. (\ref{HI}). Therefore, from the identifications (\ref{int-H-t-Identify}), we
easily find
\begin{eqnarray}\label{PS-ht-pol}
\Delta'^{(1)} \langle \bar{h^{\lambda}} \bar{t}^{\lambda}\rangle = 
\frac{2\sqrt{I \epsilon}}{k^3\tau_{e}^2}  \Big[ N_{e} - \frac{(\e^2/\epsilon)}{63 H^2} \Big]\,,
\end{eqnarray}
which can also be justified from the direct in-in calculation.

Having obtained the  two point function of ${\bar h}^\lambda$ and ${\bar t}^\lambda$ and their cross-correlation we can obtain the power spectra. The power spectrum of the gravitational tensor modes as usual are defined via 
\begin{eqnarray}
\label{PS-hh}
\sum_{+,\times}{ 
\big\langle { { h}^{\lambda}}^\dagger(\tau,{\bf k})\, { { h}^{\lambda'}}(\tau,{\bf k}') \big\rangle} = 
2{\langle {h^\lambda}^\dagger {h^{\lambda}} \rangle} \,  
\equiv \frac{2 \pi^2}{k^{3}}  \mathcal{P}_{ h} \, (2 \pi)^3\delta^{(3)}({\bf k}-{\bf k}') \,.
\end{eqnarray}
From Eq. (\ref{tensor-canonical}) we have ${\langle {h^\lambda}^\dagger {h^{\lambda}} \rangle} 
= \frac{4{\langle {\overline h^\lambda}^\dagger {\overline h^{\lambda}} \rangle}}{a^2}$,  which after substituting from Eq. (\ref{hh-correction}), we obtain the following expression for the power spectrum of the gravitational  tensor modes
\begin{eqnarray}
\label{PS-hh-f}
\mathcal{P}_{h} = \mathcal{P}_{h}^{(0)} \Big(1+16 I \epsilon N_{e}^{2}F(\hat{\beta})\Big) \,,
\end{eqnarray}
in which
\be\label{PS-hh-0}
\mathcal{P}_{h}^{(0)} \equiv \frac{2H^2}{\pi^2} \,,
\ee
is the standard tensor power spectrum for gravitons. The function $F(\hat{\beta})$ is defined as in 
Eq. (\ref{betA}) with the new dimensionless parameter $\hat{\beta}$ given in terms of $\beta$ as
\ba
\hat{\beta}\equiv \frac{2 \beta}{\epsilon} \, .
\ea
Interestingly, the corrections induced from the gauge fields dynamics in gravitational tensor power spectrum 
in Eq. (\ref{PS-hh-f}) has the same form as statistical anisotropy induced in tensor power spectrum in model of charged anisotropic inflation \cite{Chen:2014eua}. As discussed before, with $\e \gtrsim 10^{-3}$ we have $\beta \gtrsim 1$ and therefore one can easily have  $\hat \beta \gtrsim  100$.  In order for our perturbative approach to be valid, we require that $16 I \epsilon N_{e}^{2}F(\hat{\beta}) \ll1$. Using the form of the function $F(\hat{\beta})$ and the definition of $\hat \beta$, this is translated into 
\ba
\e  \lesssim   \frac{10 H}{M_P} \left(\frac{\epsilon}{16 I} \right)^{1/4} \sim 10^{-3} \, ,
\ea
in which the approximations $I \lesssim 10^{-4}, \epsilon \sim 10^{-2}$ and $H/M_P \sim 10^{-5}$ have been used to obtain the final result. In conclusion, for $\e > 10^{-3}$ or so, the corrections induced from the gauge field into the gravitational tensor power spectrum becomes large and our perturbative approximations break down. This conclusion is in line with the result obtained in \cite{Chen:2014eua}.  

Similarly, for  $\mathcal{P}_{t}$ and $ \mathcal{P}_{ht} $, we find 
\begin{eqnarray}
\label{PS-tt-f} &&\mathcal{P}_{ t} = \mathcal{P}_{h}^{(0)}
\Big[ 1+ \frac{20 }{3} I \epsilon N_e +    16 I \epsilon N_e^2 \big(1- F(\hat \beta) \big) \Big] \,, \\
\label{PS-ht-f} && \mathcal{P}_{ht}= 4\sqrt{I \epsilon} \, \mathcal{P}_{h}^{(0)} N_{e} 
\big(1- \hat{\beta} \big)\,.
\end{eqnarray}
We see interesting similarities between $\mathcal{P}_{ t} $ and $\mathcal{P}_{ S} $ in Eq. (\ref{PS}) and 
between $\mathcal{P}_{ht}$ and $\mathcal{P}_{ S{\calR}}$ in Eq. (\ref{PS-SR}).

Having calculated the curvature perturbation and the gravitational tensor power spectra in  
Eqs. (\ref{PS-R}) and (\ref{PS-hh-f}), the ratio of the tensor to scalar power spectra, denoted by the parameter $r$, is given by
\ba
\label{r-eq}
r\simeq 16 \epsilon \big( 1- 16 I N_e^2F(\beta) + 16 I \epsilon N_e^2F(\hat\beta) \big) \, .
\ea
For large enough $\hat \beta$, the last term above dominates  over the second term and we will
have a positive contribution for $r$, modifying the standard result $r = 16 \epsilon$ in single field slow-roll models of inflation.  For example, if we take $\e$ such that $\hat \beta \sim 10$, then the last term above is at the order of unity in chaotic model.   A large value of $r$ is  disfavoured  in light of the recent constraint  $r \lesssim 0.07$ \cite{Ade:2018gkx}.


\section{Summary and Conclusions}
\label{summary}

In this work we considered a model of inflation containing three complex scalar fields charged under $U(1)_a$ gauge symmetry with gauge coupling $\e$. The corresponding gauge fields $A^{(a)}_\mu$ enjoy an internal $O(3)$ symmetry associated with the rotation in field space. In a sense this model is a hybrid of models of anisotropic inflation and models based on non-Abelian gauge fields \cite{Maleknejad:2011jw, Adshead:2012kp, Maleknejad:2014wsa, Agrawal:2017awz, Agrawal:2018mrg, Maleknejad:2018nxz, Dimastrogiovanni:2018xnn}. Similar to anisotropic inflation models, with appropriate coupling of the gauge fields to the inflaton field, the system reaches an attractor phase in which the energy density of the gauge fields  reaches a constant fraction of the total energy density and the gauge field perturbations become scale invariant.

We have decomposed the scalar perturbations into the adiabatic and entropy modes. The corrections from the gauge fields into the curvature perturbations are given by Eq. (\ref{PS-R}) where the effects of gauge coupling is captured by the function $F(\beta)$.  As expected, it has the same structure as in models of anisotropic inflation, i.e. being proportional to $I N_e^2$. However, because of the background isotropy, no quadrupolar statistical anisotropy is generated.  We have also calculated the corrections in spectral index. Requiring a nearly scale invariant  curvature perturbation power spectrum requires $I \lesssim \epsilon/10 N_e \sim 10^{-4}$. This should be compared to models of anisotropic inflation in which the amplitude of quadrupolar anisotropy $g_*$ is given by $g_* = 24 I N_e^2$ and demanding $| g_*| \lesssim 10^{-2}$ from CMB observations requires $I \lesssim 10^{-7}$. 

We have calculated the tensor power spectra of the model. In addition to tensor perturbations  coming from the metric sector,  we also have new tensor perturbations from the gauge fields sector. The interactions between the matter and metric tensor perturbations induce corrections into the primordial gravitational wave spectra given by  Eq. (\ref{PS-hh-f}). We have shown that the effects of gauge coupling $\e$ are more pronounced in tensor power spectrum, controlled by the function $F(\hat \beta)$. 
For example, in simple model of chaotic inflation with $H/M_P \sim 10^{-5}$, we require $\e \lesssim 10^{-3}$ in order for the corrections in tensor power spectrum to be perturbatively under control. This is originated from the interaction  $\e^2 g^{\mu \nu} A^{(a)}_\mu A^{(a)}_\nu \phi^2$  as in Higgs mechanism. In large field model with $\phi > M_P$, large interactions between the tensor perturbations and gauge field perturbations are generated which induce large corrections in tensor power spectrum. 
We also calculated the power spectrum of the matter tensor perturbation and the cross correlation between the matter and metric tensor perturbations, given respectively by Eqs. (\ref{PS-tt-f}) and  (\ref{PS-ht-f}).

One shortcoming of our analysis is that in order to simplify the setup we have restricted ourselves to the 
subset of the model where  $\phi_{(1)} = \phi_{(2)} = \phi_{(3)} \equiv  \phi/\sqrt{3}$. This  requires some levels of fine-tuning.  However, similar to the analysis of \cite{Yamamoto:2012tq},  one expects that the isotropic FRW background is an attractor solution at least in some corners of model parameters so we may assume 
$\phi_{(1)} = \phi_{(2)} = \phi_{(3)} = \phi(t)/\sqrt{3}$ at the background level.  However, to simplify the  analysis further,  we impose a  more stronger condition and assume  that these scalar fields behave similarly at the  level of perturbations, i.e.  
 $\delta\phi_{(1)} = \delta\phi_{(2)} = \delta\phi_{(3)} = \delta\phi(t,{\bf x})/\sqrt3$.   
 If we do not take this simplification into account, we will find three entropy modes  whereas in our simplified setup  studied here the three entropy modes are treated to be identical. While we expect that the structure of the main results obtained here to remain unchanged, but it is an important question to study the general case where all three entropy modes are turned on.

There are a number of directions in which the current study can be extended. One natural question is the non-Gaussianity of the model.  In particular, in models of anisotropic inflation large anisotropic non-Gaussianities are generated. Correspondingly, we expect that observable local type   non-Gaussianity to be generated in our model. In addition, there will be cross correlation between tensor-scalar-scalar correlations which may have observable implications such as for the fossil effects \cite{Dai:2013kra, Dimastrogiovanni:2014ina, Akhshik:2014bla, Dimastrogiovanni:2015pla, Emami:2015uva, Ricciardone:2016lym, Ricciardone:2017kre}.  
Another open question in our model is the reheating mechanism which is not specified. One simple mechanism, as in standard mechanism of reheating,   is that at the end of inflation 
the gauge fields simply transfer all their energies to conventional radiation i.e. photons and other degrees of freedom in Standard Model. Another option is that the gauge fields do not decay. In this case its energy density has  the form of radiation which will be  quickly diluted in subsequent expansion of the Universe.  Another open question in our setup is the roles of the entropy perturbations. This question is also linked to the previous question about the mechanism of reheating. Observationally, there are stringent constraints on entropy perturbations. Therefore, the model should not generate too much entropy perturbations. To study this question, we have to specify how the reheating mechanism works in this model and whether or not the gauge fields decay to photon, baryons etc. Finally, in this work we did not elaborate on the observational implications of the model. It is an interesting question to study the predictions of the model for the CMB temperature perturbations and polarizations. The contributions  of the entropy modes and the corrections in primordial tensor power spectrum can have interesting observational implications in the light of the Planck CMB data.

\vspace{01cm}


{\bf Acknowledgments:}  We  thank R. Crittenden, E. Gumrukcuoglu,   
E. Komatsu and J. Soda  for insightful discussions and comments.  H. F., M. A. G. and A. K.  thank the Yukawa Institute for Theoretical Physics at Kyoto University for hospitality during the YITP symposium YKIS2018a ``General Relativity -- The Next Generation --".  H.~F.\ thanks ICG and the University of Portsmouth  for kind hospitality where this work was in progress. 


\appendix

\section{The Gauge Symmetries of the Model}
\label{gauge-symmetries}

Here we study the gauge symmetries of the model in some details. 

We have three independent gauge fields $A^{(a)}_\mu$ with gauge symmetry $U(1)_a$ and therefore we should demand that the three generators $\tau_a$ of the algebra $u(1)_a$ being independent. In the matrix notation, we choose the following representation 
\be\label{generators}
\tau_1=
\left(\begin{array}{ccc}
	1 & 0 & 0 \\
	0 & 0 & 0 \\
	0 & 0 & 0
\end{array}\right)\,, \hspace{1cm}
\tau_2=
\left(\begin{array}{ccc}
	0 & 0 & 0 \\
	0 & 1 & 0 \\
	0 & 0 & 0
\end{array}\right)\,, \hspace{1cm}
\tau_3=
\left(\begin{array}{ccc}
	0 & 0 & 0 \\
	0 & 0 & 0 \\
	0 & 0 & 1
\end{array}\right)\,.
\ee
The above matrices are clearly independent, and further satisfy 
\be\label{generators-idempotent}
\tau_a \tau_b = \tau_a \delta_{ab} \,.
\ee
Moreover, the generators in Eq. (\ref{generators}) satisfy the abelian algebra
\begin{eqnarray}\label{generators-CR}
[\tau_a, \tau_b] = 0 \,.
\end{eqnarray}

The field strength tensor associated with three copies of gauge fields are given by 
\be
\label{curvature} i \, \e  {\mathbf F}_{\mu\nu} = [{\mathbf D}_{\mu},{\mathbf D}_{\nu}] \,.  
\ee
Substituting Eq. (\ref{CD}) into Eq. (\ref{curvature}) and then using Eq. (\ref{generators-CR}) we 
find
\begin{eqnarray}
F^{(a)}_{\mu\nu} = \partial_{\mu} A^{(a)}_\nu - \partial_{\nu} A^{(a)}_\mu \,,
\end{eqnarray}
where as usual ${\mathbf F}_{\mu\nu}=F^{(a)}_{\mu\nu}\tau_a$ and ${\mathbf A}_{\mu} = 
A_{\mu}^{(a)} \tau_a$. 

Due to the abelian structure (\ref{generators-CR}) of the algebra $u(1)_a$, the gauge coupling $\e$ did not appear in the above curvature tensor which confirms that we deal with three independent copies of $U(1)$ gauge fields.

The model (\ref{action01}) is invariant under the $U(1)_a$
gauge symmetry
\begin{equation}\label{GS-general}
\Phi \to \exp({i \mathbf \Lambda} ) \Phi  \,, \hspace{1cm}
{\mathbf A}_{\mu} \to {\mathbf A}_\mu - \frac{1}{\e} \partial_{\mu}{\mathbf \Lambda} \,,
\end{equation}
where ${\mathbf \Lambda}$ is a general matrix in the field space. More specifically,  the matrix ${\mathbf \Lambda}$  can be expressed in terms of the basis as ${\mathbf \Lambda} = \lambda^{(a)} \tau_a$ which, after substituting from Eq. (\ref{generators}),  takes the form ${\mathbf \Lambda} = \mbox{diag}(\lambda^1,\lambda^2,\lambda^3)$. The gauge transformations (\ref{GS-general}) then implies

\begin{equation}\label{GF-gaugeF}
\phi_{(a)} \to \exp (i \lambda^{(a)})   \phi_{(a)}  \,, \hspace{1cm}
A^{(a)}_{\mu} \to A^{(a)}_\mu - \frac{1}{\e} \partial_{\mu}\lambda^{(a)} \,.
\end{equation}

As expected, each copy of the gauge fields $A_\mu^{(a)}$ enjoys $U(1)$ gauge symmetry. To fix the $U(1)_a$ gauge freedoms, we work in the unitary gauge where the phases of the complex scalar field are set to zero and  all scalar fields $\phi_{(a)}$ are real. 

We are interested in isotropic FRW solution so let us check if this solution can be supported in our setup.  The Maxwell kinetic term in the action (\ref{action01}) takes the component form 
$F^{(a)}_{\mu \nu} F_{(a)}^{\mu \nu}$ where we have used the fact that 
$\mbox{Tr}(\tau_a\tau_b)=\delta_{ab}$ as can easily be deduced from  
Eq. (\ref{generators}).   We see that the Maxwell kinetic term enjoys an internal $O(3)$ symmetry, i.e.  it is invariant under an $O(3)$ rotation in field space  $A_\mu^{(a)} \to R^{(a)}_{(b)} A_\mu^{(b)}$ where $R^{(a)}_{(b)}$ are the components of the $O(3)$ rotation matrices. Therefore, the Maxwell term can support an isotropic FRW background solution.  On the other hand, the  the kinetic term of the scalar 
sector in unitary gauge where all $\phi_{(a)}$ are real is given by 
\begin{eqnarray}
\label{kinetic}
({\mathbf D}_{\mu}\Phi)^\dagger ({\mathbf D}^{\mu} \Phi) &=& 
\partial_{\mu}\Phi^\dagger \partial^{\mu}\Phi 
+ \e^2 \Phi^\dagger {\mathbf A}_{\mu}^\dagger {\mathbf A}_{\mu} \Phi
+ i \e \big(\partial^{\mu}\Phi^\dagger {\mathbf A}_{\mu} \Phi 
- \Phi^\dagger {\mathbf A}_{\mu}^\dagger \partial^{\mu}\Phi \big) \nonumber \\ 
&=&  {\partial_{\mu}{\phi}}_{(a) }\partial^{\mu}\phi_{(a) } 
+ \e^2 \phi_{(a)}^2 A_{\mu}^{(a) } A_{(a) }^{\mu} \,,
\end{eqnarray}
where in the second line we have substituted from Eq. (\ref{Phi}) and the summation rule
on the repeated index $a$ is understood. 

The term  $\phi_{(a)}^2 A_{\mu}^{(a) } A_{(a) }^{\mu}$ in  Eq. (\ref{kinetic}) is not invariant under  internal $O(3)$ rotation so in general an isotropic FRW background may not be supported by this model. As mentioned in the main text,  in order to obtain an  isotropic solution  we consider a subset of the model in which  $\phi_{(1)} = \phi_{(2)} = \phi_{(3)} \equiv  \phi/\sqrt{3}$ upon which  the kinetic term (\ref{kinetic}) takes the
isotropic  form \cite{Papadopoulos:2017xxx}
\begin{eqnarray}
\label{kinetic-o3}
({\mathbf D}_{\mu}\Phi)^\dagger ({\mathbf D}^{\mu} \Phi) = 
\partial_{\mu}\phi \partial^{\mu}\phi 
+ \frac{\e^2}{3}  \phi^2 A_{\mu}^{(a) } A_{(a) }^{\mu} \,.
\end{eqnarray}
Plugging this in the starting action (\ref{action01}), yields the reduced action Eq. (\ref{action}). 


\section{Quadratic action for scalar perturbations}
\label{App-action}
Here we present the quadratic action of the scalar perturbations. As discussed in the main text, we neglect the gravitational back-reactions from the non dynamical fields $(\alpha, \beta)$.

Going to the Fourier space $\delta X(\tau,x) = \int \frac{d^3k}{(2\pi)^3} \delta X_k(\tau) 
e^{i {\bf k.x}}$ and plugging the perturbations defined in Eqs. (\ref{metric-perturbations}) and (\ref{GF-perturbations}) into the action  (\ref{action}) and  performing some integration by parts, it is cumbersome but 
straightforward to show that the quadratic action for the scalar modes is given by 
\begin{eqnarray}\label{action1}
S^{(2)}&=&\int d\tau d^3 k \Big[ 
\frac{1}{2} a^2\delta \phi '^2
- \left(\frac{1}{2} a^4 V''+\frac{1}{2} a^2 A^2\e^2+\frac{a^2 k^2}{2}-\frac{3}{2} f A'^2 f''-\frac{3}{2} A'^2 f'^2\right)\delta \phi ^2 \nonumber \\ 
&+& \frac{3}{2} f^2 \text{$\delta $Q}'^2-\left(\frac{1}{2} a^2 \e^2  \phi ^2+f^2 k^2 \right)\text{$\delta $Q}^2 +\frac{1}{2}k^4 f^2 M'^2-\frac{1}{6} \e^2 k^4 a^2 \phi^2 M^2\nonumber\\
&+& f^2 k^2 U'^2- \left(\frac{1}{3} a^2 \e^2 k^2  \phi ^2
+f^2 k^4\right)U^2 
+Y^2 \left(\frac{1}{6} a^2 \e^2 k^2 \phi ^2+\frac{f^2 k^4}{2}\right)
\nonumber\\
&+& Y \left(f^2 k^2 \text{$\delta $Q}'-k^4 f^2 M'
+ 2f k^2 A' f'\delta \phi\right) + 6 f A' f' \text{$\delta $Q}' \delta \phi 
-2 a^2 A  \e^2  \phi \text{$\delta $Q} \delta \phi  \nonumber\\ 
&+& \frac{1}{3} \e^2 a^2 k^2 \left(2A\phi \delta\phi M+\phi^2\delta Q M\right)
-2 k^2  f' f A' \delta \phi M'-k^2  f^2 \text{$\delta $Q}'M'\Big] \,,
\end{eqnarray} 
where we have represented the amplitude of the Fourier modes $\delta X_k (\tau)$ with 
$\delta X(\tau)$ and a prime indicates the derivative with respect to the conformal time $\tau$. 

From the above action, we see that the mode $Y$ is non-dynamical  which can be solved from its equation of 
motion as
\ba
\label{y}
Y=-\frac{3 f \Big(2 \delta \phi A' f'
+f \left(\text{$\delta $Q}'-k^2M'\right)\Big)}{a^2 \e^2 \phi ^2+3 f^2 k^2} \,.
\ea

We can substitute the above solution into the action (\ref{action1}). Before doing this, we
note that in the denominator of (\ref{y}) we can neglect $\e^2 a^2 \phi ^2$ in comparison
with $3 f^2 k^2$. To see this, let us find time $\tau_c$ when these two terms become 
comparable
\ba
-\tau_c=\left(-\frac{\e\phi}{Hk}\right)^{\frac{1}{3c}}\left(-\tau_e\right)^{\frac{2}{3}},
\ea
The ratio of the second term compared to the first term scales as ${\e^2a^2\over f^2}\sim{e^2H^2\tau_e^{4c}\over \tau^6}$.  Hence, during early stage of inflation in which $|\tau|\gg |\tau_e|$ the second term is negligible compared to the first term. Then the effect of gauge coupling $\e$ is subdominant at this stage and the leading interactions comes from $f(\phi)^2 F^2$. However, as inflation proceeds the effect of second term becomes important and the interaction $\e^2\phi^2 A^2$ dominates only near the time of the end of inflation. 
Therefore, neglecting $a^2 \e^2 \phi ^2$ in comparison with $3 f^2 k^2$ in Eq. (\ref{y}) and then
substituting the result into the action (\ref{action1}) we find 
\begin{eqnarray}\label{actionint}
S^{(2)}&=&\int d\tau d^3 k \Big[ 
\frac{1}{2} a^2\delta \phi '^2- \Big(\frac{1}{2} a^4 V''+\frac{1}{2} a^2 A^2\e^2+\frac{1}{2} f'^2 A'^2-\frac{2}{3} \frac{\e^2 a^2 \phi^2 f'^2 A'^2}{ k^2 f^2}+\frac{a^2 k^2}{2}\nonumber \\
&& -\frac{3}{2} f A'^2 f'' \Big)\delta \phi ^2  
+ f^2\left(1+\frac{1}{6 k^2 f^2} \e^2 a^2 \phi^2\right) \text{$\delta $Q}'^2-f^2 k^2\left(1+\frac{1}{2 f^2 k^2} a^2 \e^2  \phi ^2 \right)\text{$\delta $Q}^2\nonumber\\
&& +\frac{1}{6} \e^2 k^2 a^2 \phi^2\left( M'^2- k^2  M^2\right)
+ f^2 k^2 \mathcal{U}'^2-f^2 k^4 \left(1+\frac{1}{3 f^2 k^2} a^2 \e^2\phi ^2\right)\mathcal{U}^2\nonumber\\
&&+4 f A' f'\left(1+\frac{1}{6 k^2 f^2} \e^2 a^2 \phi^2\right) \text{$\delta $Q}' \delta \phi -2 a^2 A  \e^2  \phi \text{$\delta $Q} \delta \phi 
\nonumber\\
&&+\frac{ \e^2}{3}a^2\phi k^2 \left(2A\delta\phi M+\phi\delta Q M\right)
-\frac{2}{3 f}\e^2 a^2\phi^2  f' \mathcal{A}' \delta \phi M'-\frac{1}{3 }\e^2a^2\phi^2\text{$\delta $Q}'M'\Big] \,.
\end{eqnarray} 

We now consider the field redefinition $\bar{M}=k^2 M-\delta Q$ in terms of which the 
above action takes the following form
\begin{eqnarray}\label{AA1}
S^{(2)}&=&\int d\tau d^3 k \Big[ 
\frac{1}{2} a^2\delta \phi '^2- \Big(\frac{1}{2} a^4 V''+\frac{1}{2} a^2 A^2\e^2+\frac{1}{2} f'^2 A'^2-\frac{2}{3} \frac{\e^2 a^2 \phi^2 f'^2 A'^2}{ k^2 f^2}+\frac{a^2 k^2}{2}\nonumber \\
&& -\frac{3}{2} f A'^2 f'' \Big)\delta \phi ^2  
+ f^2 \text{$\delta $Q}'^2-f^2 k^2\left(1+\frac{1}{3 f^2 k^2} a^2 \e^2  \phi ^2 \right)\text{$\delta $Q}^2\nonumber\\
&& +\frac{1}{6 k^2} \e^2  a^2 \phi^2\left( \bar{M}'^2-   k^2 \bar{M}^2\right)
+ f^2 k^2 U'^2-f^2 k^4 \left(1+\frac{1}{3 f^2 k^2} a^2 \e^2\phi ^2\right)U^2\nonumber\\
&&+4 f A' f' \text{$\delta $Q}' \delta \phi -\frac{4}{3} a^2 A  \e^2  \phi \text{$\delta $Q} \delta \phi 
+\frac{2 \e^2}{3}a^2\phi  A\delta\phi \bar{M}
-\frac{2 \e^2}{3 k^2 f} a^2\phi^2  f' A' \delta \phi \bar{M}'\Big] .
\end{eqnarray}

The advantages of working with $\bar{M}$ is that not only the quadratic action takes a more
simple form but also  that this mode is heavy during most of the inflationary
era and we can therefore neglect it. To see this, we compare the two scalar modes 
$\delta Q$ and $\bar{M}$ in the above action as
\ba
\frac{L_{\delta Q^2}}{L_{\bar{M}^2}}\sim \frac{k^2f^2}{\e^2a^2\phi^2}\gg 1 \,,
\ea
which clearly shows that the contribution from the mode ${\bar M}$ is negligible during much of the 
period of inflation. 

Now, neglecting the subleading slow-roll corrections containing $\epsilon$ and its derivative and working to linear order in $I$ we obtain the action (\ref{action-canonical}). In principle we could calculate the 
quadratic action non-perturbatively in terms of the parameter $I$ (i.e. to all orders in powers of $I$). However, as  demonstrated in  subsection \ref{curvature-power}, requiring a nearly scale invariant corrections from the  gauge field into curvature perturbation power spectrum requires $I \ll 1$, justifying our approximation in keeping only terms linear in $I$ in quadratic action (\ref{action-canonical}). 

 In obtaining the action (\ref{action-canonical}), we have used the following formula
\ba
V &\simeq& 3 H^2 \big( 1 - \frac{\epsilon}{6} (I+2) \big) \, , \\
A' &=&\sqrt{I \epsilon} (-\tau)^{-1} \frac{a}{f} \, , \quad \quad 
\e \phi A = \e\frac{\sqrt{2I}}{3} \frac{a}{f}  \, , \\
f&=&(\tau/\tau_{e})^2 \, , \\
\phi&=& \sqrt{2/\epsilon} \, .
\ea

{}

\end{document}